\documentclass[aps,prb,twocolumn,amsmath,amssymb,floatfix,superscriptaddress]{revtex4-2}
\usepackage{blindtext}
\usepackage{epsfig}
\usepackage{color}
\usepackage{amsmath}
\usepackage{amsfonts}
\usepackage{amssymb}
\usepackage{graphicx}%
\usepackage{url}
\usepackage[breaklinks]{hyperref}
\usepackage[english]{babel}
\usepackage{esvect}
\usepackage[normalem]{ulem}
\usepackage[inline]{enumitem}
\usepackage{xcolor}
\usepackage{xfrac}
\usepackage{dsfont}
\usepackage{bm}
\usepackage{array}
\newcolumntype{C}[1]{>{\centering\let\newline\\\arraybackslash\hspace{0pt}}m{#1}}
\usepackage{footnote}
\usepackage{isomath}
\DeclareMathAlphabet\mathbfcal{OMS}{cmsy}{b}{n}

\hypersetup{colorlinks,citecolor=blue,filecolor=blue,linkcolor=blue,urlcolor=blue}

\bibpunct{[}{]}{,}{n}{}{}

\begin{document}

\title{Geometrical approach and topological electron density in the $p_x + ip_y$ superconductor}
\author{Karyn Le Hur} 
\affiliation{CPHT, CNRS, Institut Polytechnique de Paris, Route de Saclay, 91128 Palaiseau, France}
\author{Kabir Narayanan}
\affiliation{CPHT, CNRS, Institut Polytechnique de Paris, Route de Saclay, 91128 Palaiseau, France}
\affiliation{Universit\' e Paris-Saclay, CNRS, Master Quarmen, Rue Louis de Broglie, 91400 Orsay}
\author{\' Oscar Ma$\tilde{\hbox{n}}$as Chambert}
\affiliation{CPHT, CNRS, Institut Polytechnique de Paris, Route de Saclay, 91128 Palaiseau, France}
\affiliation{Laboratoire Kastler Brossel, Coll\` ege de France, CNRS, Universit\' e PSL, 11 Place Marcelin Berthelot, 75005 Paris, France}

\begin{abstract}
We present an analysis on the geometrical and physical nature of the $p_x + ip_y$ superconductor
on the square lattice, with an emphasis on the topological phase transition
at half-filling. We develop a local topological marker from specific Dirac points within the
Brillouin zone, which is introduced via the addition of two one-dimensional (1D) $\mathbb{Z}$ $(\mathbb{Z}_2$)
invariants defined on the Bloch sphere. We relate this topological marker to the electron spectral function
integrated on frequency through the local momentum-resolved electron density, which may be measured via Angle Resolved Photoemission
Spectroscopy (ARPES), and show that it remains well-protected including temperature effects. 
Integrating on a small area around a specific point in momentum space associated to the measure uncertainty, this also reveals the Van Hove logarithmic profile of the density of states in the derivative of the local marker
while preserving the topological information. Topological transitions correspond to a protected semi-metal. We analyse the real space representation of this topological marker from correlation functions. We 
present physical responses such as the topological superfluid density.
\end{abstract}
\maketitle

Topological systems are characterized through an interesting bulk-edge correspondence such that a global topological invariant reveals the protected transport at the edges. A paradigmatic situation is the quantum Hall effect \cite{KlitzingDordaPepper}. Measuring directly the invariant from the bulk is important to understand precisely what are the topological properties of a material. But, measuring directly such an invariant is generally a challenging task.  For the quantum
Hall and quantum anomalous Hall effects \cite{Haldane,QiZhang}, various methods are designed and innovated to access the geometrical properties from the bulk such as the quantum Hall conductivity \cite{TKNN}, quantized response from circular polarized light \cite{Nathan,Hamburg}. Circular dichroism for ARPES allows us to access the Berry curvature \cite{Berry} and orbital polarization in momentum space \cite{ARPES}. Measuring the topological invariant of topological superconducting systems such as the $p_x+ip_y$ superconductor \cite{ReadGreen} is yet challenging related to the Majorana fermion at the edges. This is precisely the goal of this article to propose a measurable topological invariant for the $p_x+ip_y$ superconductor. 
It is important to mention here efforts to engineer the chiral topological $p_x+ip_y$ superconductor \cite{Articlewires,Rey,Rashba}, which is an important model in physics.

Related to our quest, in 2012, Wang and Zhang \cite{WangZhang} propose to relate properties of the electron Green's function at zero frequency with some geometrical properties at a topological quantum phase transition. Recently, one of us introduces a theoretical approach allowing us to rewrite the global topological invariant in terms of the Berry gauge potentials, Berry phases or equivalently $\pi$-winding numbers at two points within the Brillouin zone corresponding to the two poles of a sphere \cite{KarynReview}. For the honeycomb Haldane model, this approach allows us to access the quantum Hall response through the local responses at Dirac points from circularly polarized light \cite{KarynReview,KarynAndrea,KarynLight,PhilippAdolfoKaryn}. A gravitational analogue with an electric field along the polar direction reproduces the quantum Hall response \cite{KarynReview,JoelKaryn}. The pseudo-spin response corresponding to the relative occupancy on the two sublattices also reveals the topological information \cite{Pachos,KarynAndrea} related to the sign of mass \cite{Haldane,NathanArticle}. This approach \cite{KarynReview} gives a physical and geometrical understandings of fractional topological numbers related to 2D topological nodal ring semimetals associated to the quantum anomalous Hall and quantum spin Hall semimetals \cite{JoelKaryn,Karynonehalf,KarynSariah1,KarynSariah2,Karynsemimetal}.  For 2D topological insulators \cite{KaneMele}, circularly polarized light is also able to rebuild the $\mathbb{Z}_2$ topological invariant from the same Dirac points \cite{KarynLight}. For the Kitaev p-wave superconducting wire \cite{Kitaev}, signatures of the topological properties are also measurable through a cavity \cite{Olesia} and through classical light via the geometrical approach \cite{FrederickKaryn}.
Here, the quest is precisely to generalize the method from the Skyrmion physics associated to a radial magnetic field on the sphere \cite{KarynReview}, to 2D topological superconductors on the square lattice. This plan will be associated to the electron spectral function {\it integrated on frequency} at specific points within the Brillouin zone $\Gamma=(0,0)$, $M=(\pi,\pi)$, $Y=(0,\pi)$ and $X=(\pi,0)$ through the Anderson pseudo-spin \cite{Anderson,TsujiAoki} in the Nambu representation and the derivation of a few formulae, which will reveal the 2D topological invariant through 1D invariants around physical observables and light i.e. ARPES. This crystallographic approach, that will also allow us to write the 2D $\mathbb{Z}$ topological invariant in terms of signs of kinetic terms at those specific points within the Brillouin zone, will then link with the Pfaffian description for the $\mathbb{Z}_2$ topological invariant of the 1D Kitaev p-wave superconductor \cite{Kitaev} and with the physics of $\mathbb{Z}_2$ topological insulators \cite{FuKane,Morice}.

In momentum space, the Hamiltonian takes the form  \begin{equation}
 H=\frac{1}{2}\sum_{{\bf k}} \psi^{\dagger}_{\bf k}
  \left(
\begin{matrix}
\xi_{\bf k} & \Delta_{\bf k}\\
\Delta_{\bf k}^* & -\xi_{\bf k}
\end{matrix}
\right)\psi_{\bf k} = \sum_{\bf k}\psi^{\dagger}_{\bf k} H({\bf k}) \psi_{\bf k}
\label{Hk}
 \end{equation}
where
\begin{eqnarray}
\xi_{\bf k} &=& -2t(\cos k_x+\cos k_y)-\mu \\ \nonumber
\Delta_{\bf k} &=& 2\Delta(\sin k_x+e^{-i\zeta}\sin k_y). 
\label{equationmodel}
\end{eqnarray}
We introduce the Nambu spinor $\psi^{\dagger}_{\bf k}=(c^{\dagger}_{\bf k}, c_{-{\bf k}})$ in the representation of units vectors $(1,0)$ and $(0,1)$ projecting on the {\it electron} and {\it hole} sectors.
The kinetic term $\xi_{\bf k}$ associated to the nearest neighbor hopping term $t$ is invariant under parity and the pairing function $\Delta_{\bf k}$ is odd under parity. 
The Planck constant $\hbar$ is fixed to one such that wave-vector components $k_i$ also measure momentum components $p_i$. In Eqs. (\ref{equationmodel}), we also fix the lattice spacing to unity.
For simplicity, we have shifted $\Delta\rightarrow -i\Delta$ and $\Delta$ is then introduced as $|\Delta|e^{i\varphi}$. We allow for a general superfuid phase shift of $\zeta$ between the two 1D superfluid order parameters.
 The $p_x+ip_y$ superconductor then will refer to the particular choice $\zeta=\frac{\pi}{2}$.
We remind the ground-state BCS wavefunction associated to the Hamiltonian
\begin{equation}
|BCS\rangle = \prod_{{\bf k}}^{\prime} (u_{\bf k}+v_{\bf k} c^{\dagger}_{\bf k} c^{\dagger}_{-{\bf k}})|0\rangle,
\end{equation}
where $\prime$ counts the presence of each Cooper pair once (and the special points $\Gamma$, $X$, $Y$ and $M$ for which $\Delta_{\bf k}$ go to zero will be addressed below).
The factors $u_{\bf k}$ and $v_{\bf k}$ take the usual forms such that $|u_{\bf k}|^2=\frac{1}{2}\left(1+\frac{\xi_{\bf k}}{E_{\bf k}}\right)$ and $|v_{\bf k}|^2=\frac{1}{2}\left(1-\frac{\xi_{\bf k}}{E_{\bf k}}\right)$ 
with $E_{\bf k}=\sqrt{\xi_{\bf k}^2 + |\Delta_{\bf k}|^2}$. The eigenvalues of the matrix defining $H({\bf k})$ are then introduced as $\pm \frac{E_{\bf k}}{2}$.

\begin{figure}[t]
\includegraphics[width=4cm]{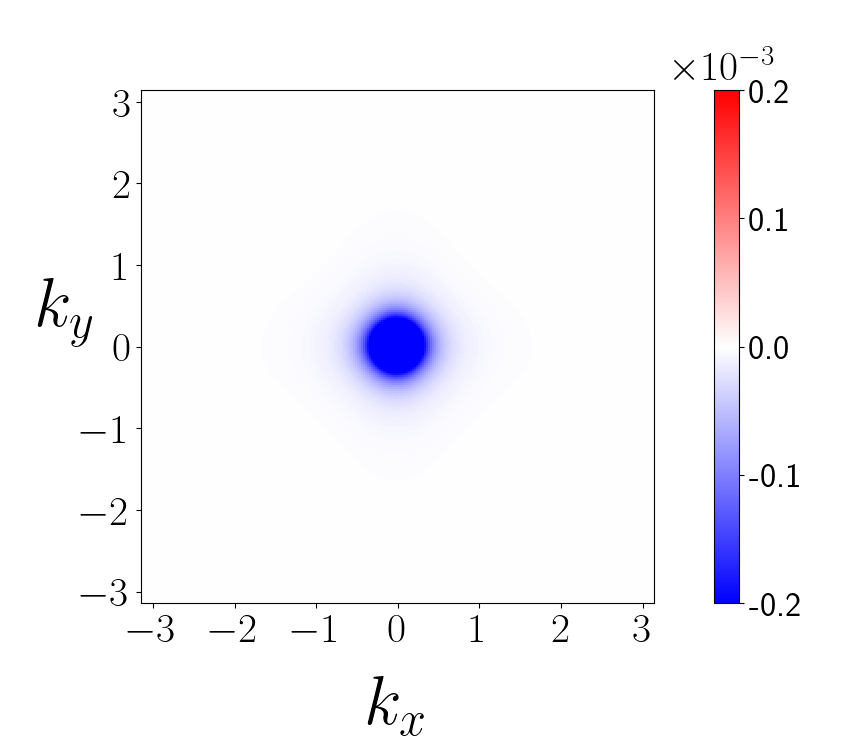}
\includegraphics[width=4cm]{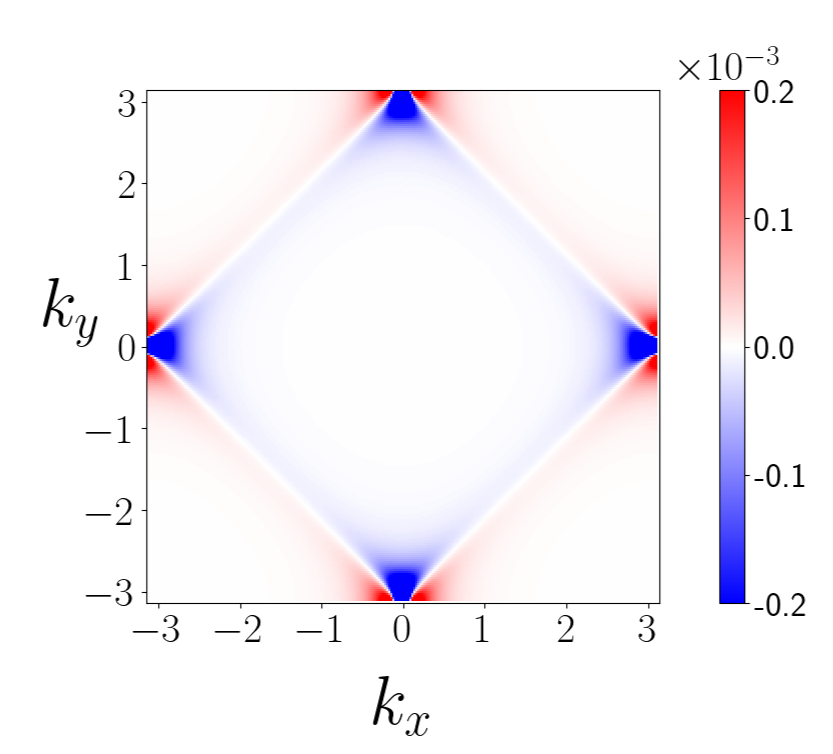}
\caption{Distribution of the Berry curvature for the $p_x+ip_y$ superconductor close to two topological phase transitions i.e. at $\Gamma=(0,0)$ for $\mu=-4t+\epsilon$ and around the $X,Y$ points when $\mu=0\pm \epsilon$ with $\epsilon\rightarrow 0$. 
Here, $\Delta=0.2$ in units of $t$.}
\label{BerrycurvaturesFig}
\end{figure}

To provide a first physical justification of the proposed local topological invariant, we begin with the profile of the Berry curvature \cite{Berry} in  Fig. \ref{BerrycurvaturesFig} obtained with the method in Ref. \cite{HatsugaiFukui}.
This will lead to equivalent definitions of the local topological marker in Eqs. (\ref{eq1}), (\ref{eq2}), (\ref{topo1}) and (\ref{topo2}) revealing the three topological quantum phase transitions in the model and the physical electron density at the specific points 
$\Gamma$, $X$, $Y$ and $M$ in Eq. (\ref{density}). 
 There is a quantum transition at $\mu=-4t$ associated to the gap closure at the $\Gamma=(0,0)$ point in reciprocal space \cite{ReadGreen}. Close to $\Gamma=(0,0)$, introducing $m=(-4t-\mu)\rightarrow 0$, we find the eigenstates to linear order in $k_x$ and $k_y$ from which the Berry curvature is evaluated (see \cite{SM}): $F_{k_x k_y}(\Gamma) =\frac{2|\Delta|^2}{m^2}\hbox{sgn}(m) \sin\zeta$.
The limit $m\rightarrow 0$ should be thought of more generally as $\xi_{\bf k}\rightarrow 0$. This result can equivalently be found through the approach of the ${\bf d}$-vector \cite{WangZhang}. This method refers to the introduction of a spin-1/2 model through the $2\times 2$ matrix where ${\bf d}$ refers to the effective magnetic field \cite{SM}. $F_{k_x k_y}(\Gamma)$ is elegant because it reveals nicely the $\hbox{sgn}(m)$ (jump-type) response
around a gapless Dirac point at a topological phase transition. This allows us to justify the introduction of the integrated Berry curvature $\tilde{F}_{k_x k_y}(\Gamma)=\frac{1}{2}\hbox{sgn}(m)\sin(\zeta)$ on the radius 
$R\sim \left|\frac{m}{\Delta}\right|\frac{1}{\sin\zeta}$ \cite{SM}. Since we fix the lattice spacing to one the momentum components are dimensionless and similarly for $R$.
Here, $\zeta\neq (0,\pi)$ such that the material is topological and $\zeta=\frac{\pi}{2}$ for the $p_x+i p_y$ superconductor. The form of the integrated Berry curvature is robust when adding a small off-diagonal hopping term $t'$ or long-range (real) hopping terms. 
A small $t'$ would shift the role of $\mu$ at each transition \cite{SM}. 

\begin{figure}[t]
\includegraphics[width=9cm]{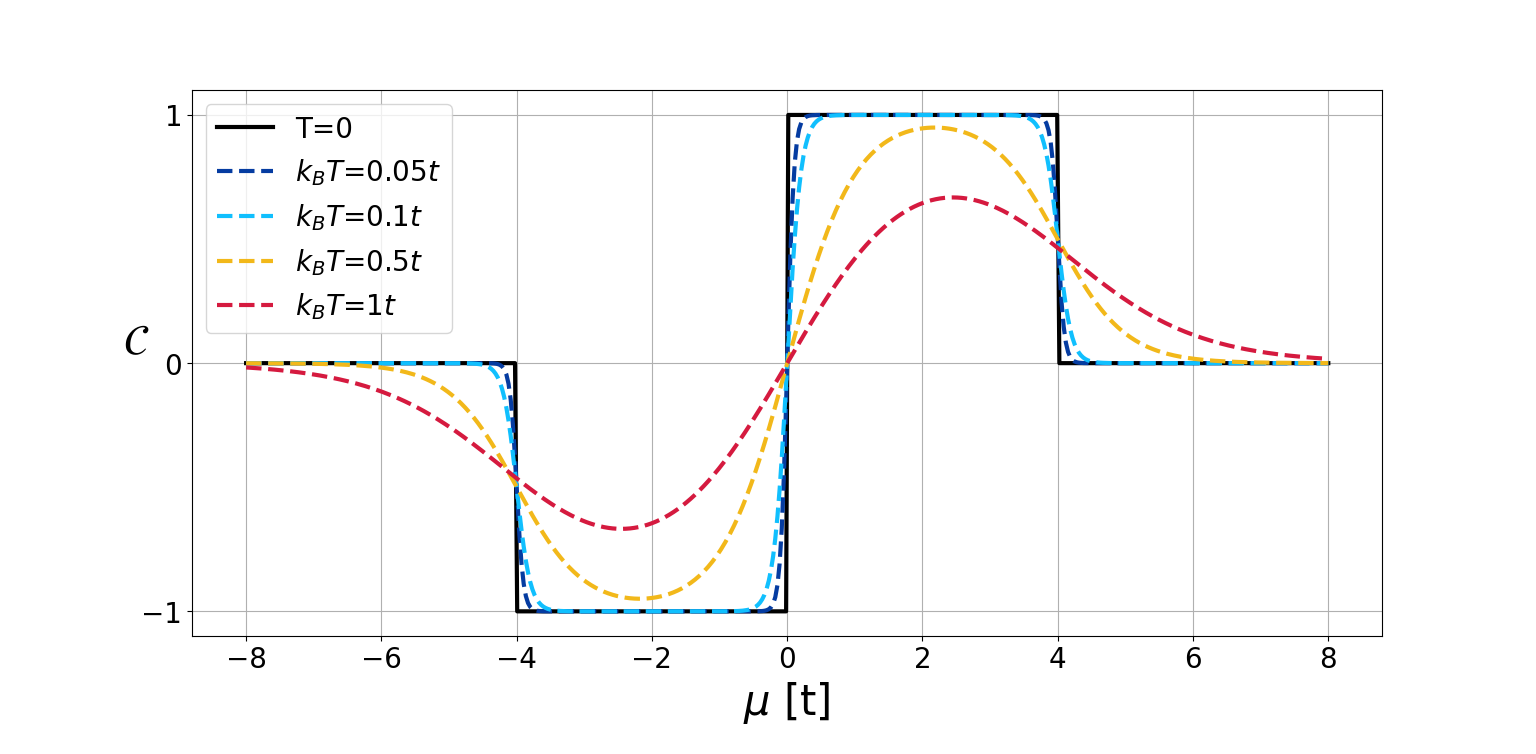}
\caption{Topological marker in Eq. (\ref{eq2}) as a function of $\mu$ for different temperatures in units of $t=1$ with $k_B=1$.}
\label{FigPhaseDiagram}
\end{figure}

We will then add geometrical properties at the $M$, $X$, $Y$ points dominating the responses at different chemical potential values.  An interesting result is that when summing the different results then this will precisely reproduce the topological
invariant obtained when integrating the Berry curvature directly on the whole Brillouin zone for each value of $\mu$. Geometrical properties at the points $M=(\pi,\pi)$ and $X=(0,\pi), Y=(\pi,0)$ can be obtained in a similar way adjusting the parameters. In particular, the transition at half-filling $\mu=0$ is dominated from the Berry curvatures centered at $(\pi,0)$ and $(0,\pi)$. See Fig. \ref{BerrycurvaturesFig}. Around $(\pi,0)$, we introduce $m=-\mu$, $k_x\rightarrow \pi-\delta k_x$ and $k_y\rightarrow k_y$ with $(|\delta k_x|,|k_y|\ll 1)$ such that $F_{\delta k_x k_y}(X=(\pi,0)) = + \frac{2|\Delta|^2}{m^2}\hbox{sgn}(\mu)\sin \zeta$.
We can then proceed as around $(0,0)$ and regularize the integral with a small disk of radius $\sim \left|\frac{m}{\Delta}\right|$. The Berry curvature around $\mu=0$ is equally distributed around the points $(\pm \pi,0)$ and $(0,\pm \pi)$. 
The points with relative $\pm$ signs allow us to rebuild a whole disk from two half disks where the Berry  curvature develops (see Fig. \ref{BerrycurvaturesFig}).
The weight of the integrated Berry curvatures is additive such that $\tilde{F}_{\delta k_x k_y}(X)+\tilde{F}_{\delta k_x k_y}(Y)=\hbox{sgn}(\mu)\sin\zeta$.
The Berry curvature located around $M=(\pi,\pi)$ can be found in a similar way with $m=4t-\mu$, $k_x\rightarrow \pi-\delta k_x$ and $k_y\rightarrow \pi-\delta k_y$ such that 
$\tilde{F}_{\delta k_x k_y}(M=(\pi,\pi))=\frac{1}{2}\hbox{sgn}(4t-\mu)\sin\zeta$. The $M$ point is important e.g. to describe the topological phase transition at $\mu=+4t$.
The integrated Berry curvatures also reveal the spread of the delocalization of Berry curvature through the increasing radius $\sim \left|\frac{m}{\Delta}\right|$,
 from one point $(\Gamma,M,X,Y)$ to another when varying $\mu$. For a fixed $\mu$, then we introduce the sum of the integrated Berry curvatures as a local marker (with $\zeta=\frac{\pi}{2}$),
\begin{eqnarray}
\label{eq1}
\Delta S =  \left(-\frac{1}{2}\hbox{sgn}(4t+\mu) + \hbox{sgn}(\mu) + \frac{1}{2}\hbox{sgn}(4t-\mu)\right).
\end{eqnarray}
The local marker $\Delta S$ reproduces the topological phase diagram in Fig. \ref{FigPhaseDiagram}. We verify that (1D) Majorana fermions do exist along an edge \cite{SM}. 

In the Haldane model \cite{Haldane}, the global topological invariant can also be re-built from local Berry curvatures around Dirac points and the $M$ point(s) \cite{KarynAndrea}. A similar phase diagram can occur for Chern insulators \cite{Bernevig}.

It is then relevant to relate topological properties with observables such as the local electron density that will question if some topological aspects may be found within the Luttinger theorem or the Fermi function itself i.e.
in the way we fill the electron states on the Fermi surface corresponding here to the time-reversal invariant points where $\Delta_{\bf k}\rightarrow 0$. 
For topological quantum Hall systems, a relation is found between the Luttinger theorem and 
many-body Chern number through the Streda formula \cite{Lucila}. We introduce the electron density from the electron spectral function for the situation of superconductivity. 
The retarded Green's function reads
\begin{equation}
G_{\bf k}(\omega) = \frac{|u_{\bf k}|^2}{\omega-E_{\bf k}+i0^+}+\frac{|v_{\bf k}|^2}{\omega+E_{\bf k}+i0^+}
\end{equation}
such that the electron density in momentum space can be measured from the electron spectral function $A({\bf k},\omega)=-\frac{1}{\pi}\hbox{Im} G_{\bf k}(\omega)$ through ARPES, as follows
\begin{equation}
\langle c^{\dagger}_{\bf k} c_{\bf k}\rangle = \langle BCS|c^{\dagger}_{\bf k} c_{\bf k}|BCS\rangle = \int d\omega n_F(\omega) A({\bf k},\omega).
\label{densityspectral}
\end{equation}
Here, $n_F$ is the Fermi factor at a temperature $T$ or inverse temperature $\beta$. At zero temperature, we can verify the result $\langle c^{\dagger}_{\bf k} c_{\bf k}\rangle =|v_{\bf k}|^2$. 
For specific points in the Brillouin zone where $\Delta_{\bf k}\rightarrow 0$, including the $\Gamma$, $M$, $X$ and $Y$ points,
\begin{equation}
\label{density}
\langle c^{\dagger}_{\bf k} c_{\bf k}\rangle_{\Delta_{\bf k}\rightarrow 0} = \frac{1}{2} - \frac{1}{2}\hbox{sgn}(\xi_{\bf k})\tanh\left(\frac{\beta E_{\bf k}}{2}\right).
\end{equation}
The local electron density then becomes topological in the sense that the topological marker can be written as
\begin{eqnarray}
\label{eq2}
\Delta S = C = -\langle c^{\dagger}_{\Gamma} c_{\Gamma}\rangle +2\langle c^{\dagger}_X c_X\rangle - \langle c^{\dagger}_M c_M\rangle.
\end{eqnarray} 
Through Eq. (\ref{density}), we find a first interesting relation with the $\hbox{sign}$ of kinetic terms at those specific points.

Eq. (\ref{eq2}) allows us e.g. to address the stability of topological properties at finite temperature through Eq. (\ref{density}). 
A plateau with a quantized invariant in Fig. \ref{FigPhaseDiagram} would remain visible even when $T\sim 10^3K$ if we set $t\sim 1\hbox{eV}$. 
In reality, the topological invariant is protected by the superconducting gap which also implies that the temperature should be smaller than the critical temperature $T_c$. 
The topological phase diagram is also protected when taking into account the ``resolution'' of the measure with light in momentum space \cite{SM}, which shows optimism for experimental applications related to the electron 
occupancy in momentum space. Integrating on a small area $\sim {\cal R}^2$ the electron density at a $X$-point, we also show that this turns the $\delta$ (step) function in the derivative of Eq. (\ref{density}) 
onto a logarithmic profile which then reveals the density of states when approaching the (small) Van Hove filling. More precisely, when the chemical potential satisfies $\mu\sim 2\frac{|\Delta|^2}{t}$, with e.g. $|\Delta|\sim 10K$ and $t\sim 1eV$, 
assuming that ${\cal R}\gg \frac{\mu}{\alpha}$ where $\alpha=\sqrt{4|\Delta|^2+2\mu t}\sim 2\sqrt{2}|\Delta|$, the integrated density around the $X$ or $Y$ point satisfies 
\begin{equation}
\frac{1}{{\cal R}^2}\iint^{{\cal R}^2} d^2{\bf k}\langle c^{\dagger}_{\bf k} c_{\bf k}\rangle \sim \frac{1}{2} - \frac{\mu}{2{\cal R}\alpha}\ln\frac{|\mu|}{{\cal R}\alpha}.
\end{equation}
A finite (small) lifetime for the quasiparticles or an integration on a specific range of frequencies in Eq. (\ref{densityspectral}) yet leads to the same topological invariant \cite{SM}. 

\begin{figure}[t]
\hskip 3cm \includegraphics[width=8cm]{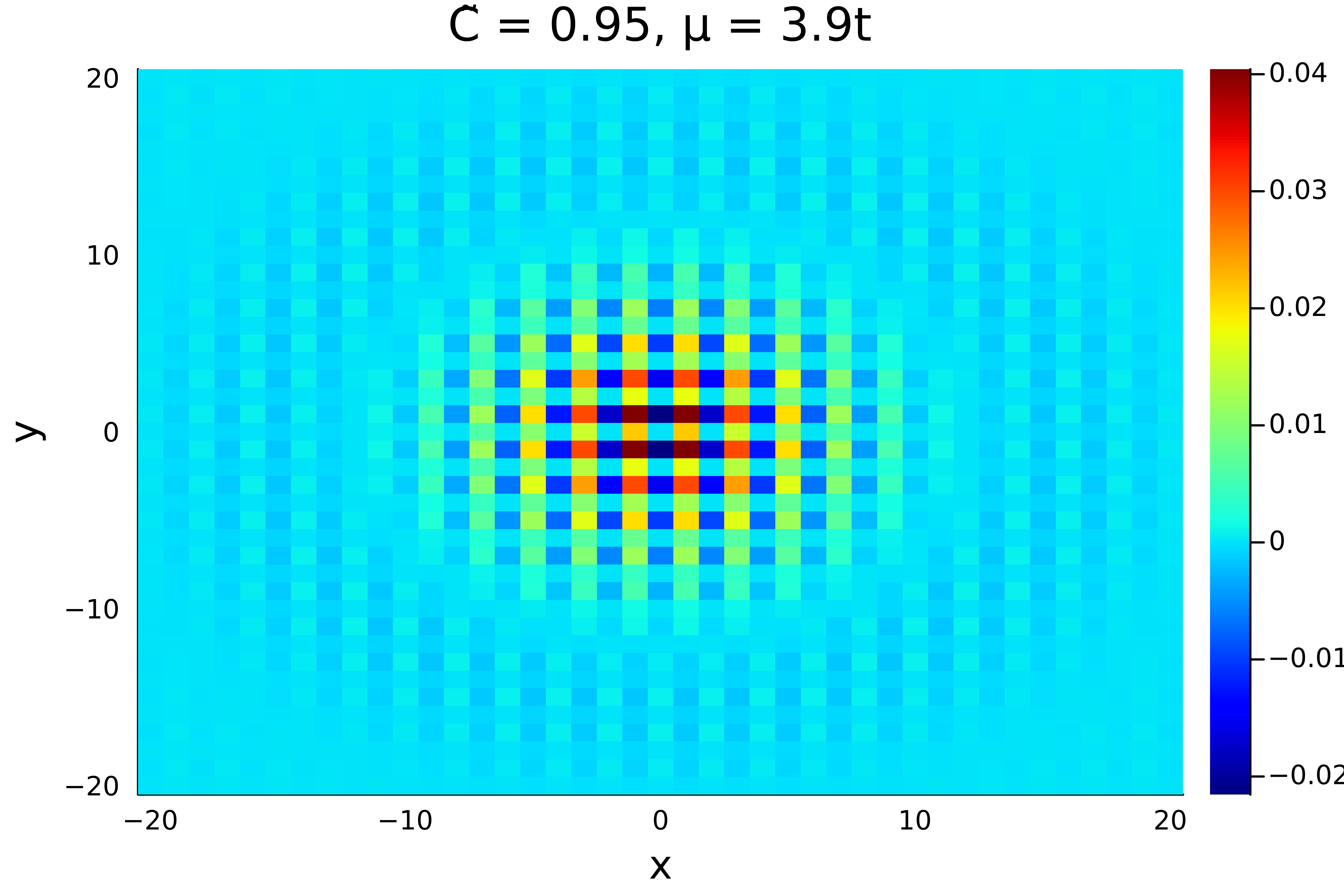}
\caption{Cartography of the local marker through $\langle c^{\dagger}_i c_j\rangle (- 1 +2 e^{i {\bf k}_X \cdot ({\bf r}_i - {\bf r}_j)} - e^{i {\bf k}_M \cdot ({\bf r}_i - {\bf r}_j)})$ for $\mu=3.9t$. 
We have fixed $\Delta=0.1$ in units of $t$. Eq. (\ref{Cequation}) in real space then allows us to identify an almost quantized topological number. 
}
\label{TopologicalMarker}
\end{figure}

From Fourier transform, we can also propose a real-space analysis of  Eq. (\ref{eq2}) via correlation functions:
\begin{equation}
C = \frac{1}{N}\sum_{i,j} \langle c^{\dagger}_i c_j\rangle (- 1 +2 e^{i {\bf k}_X \cdot ({\bf r}_i - {\bf r}_j)} - e^{i {\bf k}_M \cdot ({\bf r}_i - {\bf r}_j)}),
\label{Cequation}
\end{equation}
with $N$ the number of sites.
Introducing $({\bf r}_i - {\bf r}_j)=(\Delta_x,\Delta_y)$, the correlation functions read (for $i\neq j$) \cite{SM}
 \begin{eqnarray}
\hskip -0.6cm \langle c^{\dagger}_i c_j\rangle = \frac{-1}{2\pi^2}\int_{0}^{\pi}\int_{0}^{\pi} dk_x dk_y \cos(k_x \Delta_x) \cos(k_y \Delta_y)\frac{\xi_{\bf k}}{E_{\bf k}}.
\label{behaviorcorrelations}
 \end{eqnarray}
 At $\mu=0$, due to particle-hole symmetry we find $\langle c^{\dagger}_i c_j\rangle=0$ along the direction $\Delta_x=\Delta_y$ if $i\neq j$. This is a signature of the phase transition at $\mu=0$.
 At $\mu=0$, at long distances, associated  to specific directions in real space \cite{SM}, we also find a pure long-range algebraic decay. A power-law decay also 
 occurs at 1D topological quantum phase transitions \cite{FrederickLoicKaryn,FrederickLoicOlesiaKaryn,KarynFanMagali}. In the analysis of Read and Green \cite{ReadGreen}, through an analogy to the Pfaffian,
 a power-law behavior for the Fourier transform of $g_{\bf k}=\frac{v_{\bf k}}{u_{\bf k}}$ was reported in the weak-paired phase.
Close to $\mu=-4t$ and $4t$, within the topological phases, numerically we identify an exponential multiplying factor accompanying the power-law behavior in $\frac{1}{\sqrt{r}}$ at long distance revealing then the vicinity of a strong-paired phase (for $\mu<-4t$). When increasing the chemical potential, the local density response corresponding to $\Delta_x=\Delta_y=0$ becomes important approaching $1$ when $\mu\rightarrow +4t$ (see \cite{SM}). This central point (response) at $\Delta_x=\Delta_y=0$ gives also a zero response in the formula for $C$ in Eq. (\ref{Cequation}) in real space. Then, this requires to include the participation of long-range correlation functions of much weaker intensity to re-build the quantized topological number. The formula
 in real space in Eq. (\ref{Cequation}) is yet efficient in this case; see Fig. \ref{TopologicalMarker}. 
 The derivative of the local electron density in real space reveals a logarithmic behavior as a function of $\mu$,
 which should be detectable in the local compressibility for a wide domain in $\mu$ around half-filling \cite{SM}.  This derivative of the local electron density also reveals the topological transitions at $\mu=-4t$ and $\mu=+4t$ through a jump \cite{SM}.
 
We show below that the local marker $C$ is in fact equivalent to two 1D $\mathbb{Z}$ invariants $C_x$ and $C_y$ introduced on the sphere, which will also lead to Eq. (\ref{topo1}) introducing a correspondence with 1D Pfaffians.
Suppose that we walk from the point $(0,0)$ to $(\pi,0)$, that describes well in principle the physics for negative $\mu$, when $-4t<\mu<0$,  and follow the evolution of the integrated Berry curvature. 
The $2\times 2$ matrix associated to the Hamiltonian indeed takes the form of a $p_x$ Kitaev superfluid \cite{Kitaev} fixing $k_y=0$, with the kinetic term $\xi_{k_x}=-2t\cos k_x-(\mu+2t)$
and the pairing function $\Delta_{k_x}=2\Delta\sin k_x$. The line at $k_y=0$ then will link $\Gamma=(0,0)$ with the $X=(\pi,0)$ point within the Brillouin zone.
For this 1D $p_x$ superfluid, through an analogy with a Skyrmion, we can associate a $\mathbb{Z}$ invariant on the Bloch sphere \cite{SatoAndo,FrederickLoicKaryn}
with the polar angle $\theta\in [0;\pi]$ linking the points $(0,0)$ to $(\pi,0)$ and associated to the $k_x$ variable. The azimuthal angle $\phi$ on the sphere is then related to the superfluid phase associated to $\Delta$. 
We can map the 2D space $(k_x,\varphi)$ onto $(\theta,\phi)$ in spherical coordinates
and introduce the 1D invariant \cite{SM,FrederickLoicKaryn} $C_x = \frac{1}{2}\left(\langle S_z(\theta_{(\pi,0)})\rangle - \langle S_z(\theta_{(0,0)})\rangle \right)$ such that $C_x=-1$ for $-4t<\mu<0$, 
with the pseudo-spin component $S_z=c^{\dagger}_{\bf k} c_{\bf k}-c_{-{\bf k}} c^{\dagger}_{-{\bf k}}$ and $\langle S_z\rangle = 2\langle c^{\dagger}_{\bf k} c_{\bf k}\rangle -1$. It can also be understood as a 1D integral
in momentum space
\begin{equation}
C_x = \frac{1}{2}\int_0^{\pi} dk_x\frac{\partial \langle S_z(k_x,k_y=0)\rangle}{\partial k_x}.
\end{equation}
This invariant describes the evolution of pseudo-spin from $(0,0)$ to $(\pi,0)$. But, it will remain in fact applicable for all $\mu$ because $C_x=0$ when $\mu>0$ because 
$\langle S_z(\theta_{(\pi,0)})\rangle=\langle S_z(\theta_{(0,0)})\rangle$. Introducing the superfluid phase as an additional dimension may seem artificial, however we show below that this will allow us to acquire some physical 
insight on proposing protocols. 

To account for the physics at positive $\mu$, then we invoke a 1D $p_y$ superfluid fixing $k_x=0$ with the kinetic term
 $\xi_{k_y}=-2t\cos k_y -(\mu-2t)$ i.e. with an effective chemical potential $\mu_{eff}=\mu-2t$ (instead of $\mu_{eff}=\mu+2t$ for $C_x$). 
 We navigate on a line from the $M=(\pi,\pi)$ point to the $X=(\pi,0)$ point in the Brillouin zone.
 In that case, $k_y$ is related to the polar angle $\theta$ and, fixing e.g. $\zeta=\frac{\pi}{2}$, the superfluid phase $\varphi$ is again related to the azimuthal angle $\phi$ on the sphere.
 This results in the 1D invariant \cite{SM} $C_y = \frac{1}{2}\left(\langle S_z(\theta_{(\pi,0)})\rangle - \langle S_z(\theta_{(\pi,\pi)})\rangle \right)$. This invariant describes well in particular the transition of the 2D topological invariant at $\mu_{eff}=\mu-2t=+2t$ and the delocalization of the Berry curvature from the $X$ point onto the $M=(\pi,\pi)$ point. This invariant reads $C_y=0$ when $\mu<0$. Importantly, sticking the two 1D invariants together or simply adding them, such that $C=C_x+C_y$, 
 we indeed reproduce $\Delta S$ and Eq. (\ref{eq2}). The addition of these two 1D invariants gives a geometrical interpretation to the $p_x+ip_y$ character which is yet valid at $\mu=0$. 
We could have equally navigated from $\Gamma=(0,0)$ to $Y=(\pi,0)$ and then from $M=(\pi,\pi)$ to $Y=(\pi,0)$. 

As a remark, we can also introduce two $\mathbb{Z}_2$ 1D invariants $2C_x^2-1$ and $2C_y^2-1$ \cite{KarynReview,FrederickKaryn,SM}, which are related to the Pfaffians for the 1D Kitaev model \cite{Kitaev,SatoAndo}.
This allows us to propose a formula for the $\mathbb{Z}$ invariant in 2D as the difference of the 1D Pfaffians \cite{SM}
\begin{eqnarray}
\label{topo1}
C &=& |C_y|-|C_x| \\ \nonumber
&=& -\frac{1}{2}\left(\prod_{X,M} \hbox{sgn}(\xi_{k_y}) - \prod_{X,\Gamma} \hbox{sgn}(\xi_{k_x})\right),
\end{eqnarray}
which reveals the symmetry around the phase diagram i.e. $C_x\rightarrow -C_y$ when flipping the sign of $\mu$.
Within this definition, this emphasizes that $C$ can be equal to $0,\pm 1$ and also to $\pm \frac{1}{2}$ at topological phase transitions.
This is equivalent to a simplified 1D $\mathbb{Z}$ invariant corresponding then to navigate on the diagonal from $\Gamma$ to $M$ \cite{KarynReview,SM}:
\begin{equation}
\label{topo2}
C = -\frac{\hbox{sgn}(\mu)}{2}\int_0^{\pi} dk_{\parallel}\frac{\partial}{\partial k_{\parallel}} \langle S_z(k_{\parallel})\rangle,
\end{equation}
with $k_{\parallel}=\frac{k_x+k_y}{2}$, which is interesting for applications.

Around half-filling, i.e. around $\mu=0$, we can then develop the correspondence with a topologically protected semimetal in momentum space characterized through the $\frac{1}{2}$ mean local electron density. 
For $-4t<\mu<4t$, it is instructive to re-write the  topological local marker as
\begin{equation}
C=C_x+C_y= 2\langle c^{\dagger}_X c_X\rangle -1.
\label{invariant1D}
\end{equation}
This formula can also be understood as $C_x=\langle c^{\dagger}_X c_X\rangle-1$ and $C_y=\langle c^{\dagger}_X c_X\rangle$.
When $\mu=0$, $C=0$ and we can verify from Eq. (\ref{density}) that $\langle c^{\dagger}_X c_X\rangle=\frac{1}{2}$ is protected at any temperature defining indeed a protected topological semimetal in momentum space.
The eigenstates are formed from Majorana fermions at points $X$ and $Y$ \cite{SM}.  
 
Then, this guides us to an interesting discussion on a correspondence between the sphere and a many-body local gauge potential associated to the BCS wavefunction through physical protocols. 
 The formulation from the poles \cite{KarynReview,JoelKaryn} leads to (see \cite{SM})
 \begin{eqnarray}
 C_y = A_{\phi}(\theta_{(\pi,0)}) - A_{\phi}(\theta_{(\pi,\pi)}).
 \label{Cy}
 \end{eqnarray}
 We have implicitly assumed the form of the Berry gauge potential $A_{\phi}=-i\langle \psi| \partial_{\phi}|\psi\rangle$ within the spin eigenstate corresponding to the BCS ground state. 
 Since we have a direct relation between the azimuthal angle on the sphere $\phi$ and the superfluid phase this is similar to introduce $A_{\varphi}(k_x,k_y)$.
 This is equivalent to introduce the many-body local gauge potential $A_{\varphi}({\bf k})$ acting on one electron forming the Cooper pair $(c^{\dagger}_{\bf k},c^{\dagger}_{-{\bf k}})$ through 
  the term $v_{\bf k}e^{i\varphi}c^{\dagger}_{\bf k} c^{\dagger}_{-{\bf k}}$ in the wavefunction; the superfluid phase occurs in a re-definition of fermions $c^{\dagger}_{\pm\bf k}\rightarrow e^{\frac{i\varphi}{2}}c^{\dagger}_{\pm\bf k}$.
  We can then introduce a twist of the superfluid phase attached to $|BCS\rangle=\prod_{\bf k} |BCS\rangle_{\bf k}$, such that
 \begin{eqnarray}
 \label{twist}
 A_{\varphi}(k_x,k_y) &=& (-i)\times_{\bf k}\hskip -0.1cm\langle BCS| \partial_{\varphi}|BCS\rangle_{\bf k} \\ \nonumber
&=& \langle c^{\dagger}_{\bf k} c_{\bf k}\rangle = \frac{1}{2}(\langle S_z\rangle +1).
 \end{eqnarray}
 It also shows a relation between the local twist of the superfluid phase in momentum space and the measure of the local density.
 The topological invariant $C_y$ and therefore $C$ can be re-written in terms of $A_{\varphi}$ at the points $X$ (or $Y$), $M$ and $\Gamma$ \cite{SM}. We also have the symmetry that $A_{\varphi}(\pi,0)=A_{\varphi}(0,\pi)$.
 For $-4t<\mu<4t$, we identify $C_y=A_{\varphi}(\pi,0)$ and $C_x=A_{\varphi}(\pi,0)-1$ such that $A_{\varphi}(\pi,0)=\frac{1}{2}$ at the topological transition at half-filling, leading to $C_x=-C_y=-\frac{1}{2}$ which has an understanding as half Skyrmions \cite{SM}. Multiplying this result by $(2\pi)$ is equivalent to associate a $\pi$ winding number for this semimetallic Dirac  point similarly as for the quantum anomalous Hall semimetal \cite{JoelKaryn,Karynonehalf,KarynSariah1,KarynSariah2}.
 In terms of $\phi$, this measures the $\pi$ Berry phase around the point.  
 
 It is relevant to write precisely the forms of the Berry potentials $A_{k_x}$ and $A_{k_y}$ in the plane associated to the Berry curvature $F_{k_x k_y}=\partial_{k_x}A_{k_y}-\partial_{k_y}A_{k_x}$ \cite{SM}. From the linearized eigenstate around a topological phase transition, e.g. around $\mu=-4t$, for $m>0$, 
this agrees with 
 \begin{eqnarray}
 \label{responses}
 \hbox{Re} A_{k_y} &=& \sin \zeta \frac{|\Delta|^2}{m^2} k_x \\ \nonumber
 \hbox{Re}A_{k_x} &=& -\sin \zeta \frac{|\Delta|^2}{m^2} k_y.
 \end{eqnarray}
 The classical vector potential shifts the momentum which is then related to a current density. 
 We propose then an analogy, through Eqs. (\ref{responses}), between the Berry gauge potential on the left which is the analogue of the classical vector potential \cite{Berry,KarynReview} and 
 a transverse momentum $k_x$ associated to a topological superfluid density on the right, as follows. The operator $i\partial_{k_y}$ in quantum mechanics entering in $A_{k_y}$ \cite{SM} is equivalent to a polarization along $y$ direction in real space.
 We suppose that this polarization is produced from a force along $y$ that may be e.g. created from the electromagnetic wave. From Newton equation for an electron, a Cooper pair, this will then produce 
 a small boost of momentum $\delta k_y=\delta k_y(t)$. We can also associate a topological superfluid velocity or current density along $x$ direction
 through $\sin \zeta \frac{|\Delta|^2}{m^2} k_x$ in Eq. (\ref{responses}).
 This is an analogue of the Karplus-Luttinger velocity in quantum (anomalous) Hall systems \cite{KarplusLuttinger}.
 Indeed, describing a Cooper pair wavefunction in a macroscopic way as, $\psi \sim \sqrt{n_s}e^{i\varphi({\bf r})} \sim \psi_0 e^{2i{\bf k}\cdot {\bf R}}$ with ${\bf R}=(x,y)$  the center of mass coordinate of the Cooper pair, the superfluid current density $j_x$ along $x$ direction is indeed proportional to $\partial_x\varphi \sim 2k_x$, such that $j_x=\frac{2e}{m_s} n_s \partial_x \varphi$ (with $\hbar=1$ and $m_s$ the mass of the Cooper pair).  
 We can then introduce the (dimensionless) topological superfluid density 
 \begin{equation}
 n_s = \sin \zeta \frac{|\Delta|^2}{m^2},
 \end{equation}
  that may be measured from the local Berry curvature $F_{k_x k_y}$ via ARPES similar as in topological insulators \cite{ARPES}. From the definition of $F_{k_x k_y}$, $n_s$ is gauge invariant.
 The first equation in (\ref{responses}) then corresponds to $\hbox{Re} A_{k_y} \leftrightarrow \frac{m_s}{2e}\frac{1}{2} j_x$. The factor $\frac{1}{2}$ comes from the fact that similar to $A_{\varphi}$ we can normalize $A_{k_y}$ to the response of each electron.
 We can build a similar correspondence from the second equation in Eq. (\ref{responses}) between $\hbox{Re} A_{k_x}$ and  $-\frac{m_s}{2e}\frac{1}{2} j_y$, through $n_s$. We hope that this may lead to various applications, e.g. linked to Josephson physics and energy applications.
      
To summarize, through a correspondence from Skyrmion physics, a spin-$\frac{1}{2}$ particle in a radial magnetic field and the BCS theory of superconductivity, we present a relation between local topological markers in momentum space measurable through the local electron density and ARPES i.e. light. For the $p_x+ip_y$ superconductor, this leads to an interesting characterization of the topological phase transition at half-filling associated to a protected Dirac semimetal in momentum space with a local $\frac{1}{2}$
 invariant which can be associated to the local mean electron density. We have shown the robustness of the proposed approach towards temperature, and addressed the measure in momentum and frequency space. 
 We propose physical protocols e.g. through light and this work opens perspectives on probing the topological superfluid stiffness \cite{stiffness}. The real-space analysis of the topological invariant in Eq. (\ref{Cequation}) is also applicable
 in the presence of disorder. The local topological marker is stable when including weak disorder as in the 1D p-wave SC through correlation functions \cite{FrederickLoicOlesiaKaryn} and a stochastic topological number \cite{PhilippAdolfoKaryn,FrederickLoicOlesiaKaryn}. This approach can be applied to various topological materials.
  
 K. Narayanan and O. Ma$\tilde{\hbox{n}}$as Chambert thank CPHT at Ecole Polytechnique for the support during their respective Master Thesis associated to this work. K. Narayanan also thanks University Paris-Saclay through the Program Quarmen.
 The authors acknowledge discussions with Nathan Goldman.

\onecolumngrid
\section{Local Berry Responses and Representation on the Sphere}
\label{formalism}

We begin with geometrical properties of the $p_x+e^{i\zeta}p_y$ superconductor. 
\\

{\color{blue} Local Berry curvatures.---} To fix appropriately the different signs of the topological invariant within the whole phase diagram as a function of $\mu$, we introduce the phase
$\zeta$ associated to the pairing function $\Delta_{\bf k}^* c_{\bf -k} c_{\bf k}$ with $\Delta_{\bf k}^* = \Delta(\sin k_x+ e^{i\zeta} \sin k_y)$. 
Momenta components are labelled through $(p_x,p_y)$ and are equivalent to wavevectors components $(k_x,k_y)$ when setting the Planck constant $\hbar=1$. 

To understand what is special around $\Gamma=(0,0)$ the eigenstates can be then evaluated to {\it lowest-order} in the wave-vectors $k_x$ and $k_y$
when simplifying $H({\bf k})$ as
\begin{equation}
H({\bf k})= \frac{1}{2}\left(
\begin{matrix}
m & 2\Delta(k_x+e^{-i\zeta}k_y)\\
2\Delta^*(k_x+e^{i\zeta}k_y) & -m
\label{model}
\end{matrix}
\right),
\end{equation}
with $m=(-4t-\mu)$ and we have shifted $\Delta\rightarrow -i\Delta$. The energy eigenvalues for lower and upper bands then read $E_{\pm}=\pm \frac{E_{\bf k}}{2}$ with 
$E_{\bf k}=\sqrt{m^2+4|\Delta|^2(k_x+e^{-i\zeta} k_y)(k_x+e^{i\zeta}k_y)}$ which reveals a rotationally invariant form for
the $p\pm ip$ superconductor with $\zeta=\pm\frac{\pi}{2}$. This can be simplified as
\begin{equation}
E_{\pm}=\pm \frac{|m|}{2}\left(1+\frac{2|\Delta|^2}{m^2}\left(k_x+e^{-i\zeta}k_y\right)\left(k_x+e^{i\zeta}k_y\right)\right).
\end{equation}
The energy spectrum for small ${\bf k}$-values takes the form of a quadratic spectrum when $m\neq 0$. 
It also reveals the mass inversion effect. For $m<0$ and $m>0$ the system will respectively invert the form of the eigenstate in terms of $(0,1)^T$ and $(1,0)^T$ vectors.
At $m=0$, the topological transition then gives rise to a gapless Dirac point on the Fermi surface. Here, we verify the form of the local Berry curvature at this Dirac point
when varying locally $\mu$ around $\mu=-4t$. Close to $k_x,k_y\rightarrow 0$ we can then present the specific form of the eigenstate for the lowest-energy band. 
For $m=(-4t-\mu)<0$,
\begin{equation}
\label{eigenstate1}
|\psi_-\rangle = 
\begin{pmatrix}
-1\\
(k_x+e^{i\zeta} k_y)\frac{\Delta^*}{|m|} 
\end{pmatrix}
.
\end{equation}
For $m>0$,
\begin{equation}
\label{eigenstatemu}
|\psi_-\rangle = 
\begin{pmatrix}
-\frac{\Delta}{m}(k_x+e^{-i\zeta}k_y) \\
1
\end{pmatrix}
.
\end{equation}
This leads to a jump of the local Berry curvature around $\Gamma=(0,0)$ when $m\rightarrow 0$:
\begin{eqnarray}
\label{Berryfunction}
F_{k_x k_y} &=& \partial_{k_x} A_{k_y} - \partial_{k_y} A_{k_x} \\ \nonumber
&=&\partial_{k_x}\langle \psi_-|i \partial_{k_y}|\psi_-\rangle - \partial_{k_y}\langle \psi_-|i\partial_{k_x}|\psi_-\rangle \\ \nonumber
F_{k_x k_y}(\Gamma) &=& \frac{2|\Delta|^2}{m^2}\hbox{sgn}(m) \sin\zeta.
\end{eqnarray}
For $m\sim 0$, all the geometrical properties are then described around the $\Gamma=(0,0)$ point, as verified numerically in the Letter. 
The change in Chern number at the topological transition can be deduced from the change in Berry curvature integrated on a small disk around the Dirac point. The radius of this disk then will correspond to $R\sim \left|\frac{m}{\Delta}\right|$ in the reciprocal space with  $|m|\ll |\Delta|$ when approaching the transition at $\mu=-4t$. The variable $R$ is dimensionless implying that the variable $k=|{\bf k}|$ should also be thought of as dimensionless i.e. as $ka$ where we fix the lattice spacing $a$ to unity.
The integrated Berry curvature then reads
\begin{equation}
\tilde{F}_{k_x k_y}(\Gamma)= \iint F_{k_x k_y} \frac{dk_x dk_y}{2\pi} = \int_0^R \left(k \frac{2|\Delta|^2}{m^2}\hbox{sgn}(m) \sin\zeta\right) dk.
\end{equation}
With $R^2=\left|\frac{m}{\Delta}\right|^2 \frac{1}{2\sin\zeta}$, the integrated Berry curvature then satisfies
\begin{equation}
\tilde{F}_{k_x k_y}(\Gamma)= -\frac{1}{2}\hbox{sgn}(4t+\mu). 
\end{equation}
The factor $\frac{1}{2}$ ensures that the jump will always be an integer. The definition of $R$ also shows that for $\zeta=0$ the material would not be topological resulting in $R\rightarrow +\infty$.
It is also relevant to emphasize that this result is robust to deformations of the band structure e.g. we can add a $t'$ off-diagonal hopping term that would essentially shift the role of $\mu$ at each transition. Therefore, this result is rather a general consequence of the form of the pairing form which takes a linear form close to the specific points discussed here. 
\\

This result can also be found through the approach of the ${\bf d}$-vector acting on a spin-$\frac{1}{2}$ particle through a spin Hamiltonian $H=-{\bf d}\cdot {\bf S}$ with ${\bf S}=\frac{\mathbfit{\sigma}}{2}$ associated to $H({\bf k})$ at small ${\bf k}$-values. Suppose that for simplicity $\Delta\in\mathbfit{R}$. Then,
\begin{equation}
\label{d}
{\bf d}=(d_x,d_y,d_z)=-(2\Delta k_x +2\Delta \cos \zeta k_y, 2\Delta k_y\sin \zeta, m).
\end{equation}
In this way, the equivalent formula for the Berry curvature 
$\tilde{F}_{k_x k_y}= \frac{1}{2}\epsilon^{abc} n^a \partial_{k_x} n^b \partial_{k_y} n^c$ with $n^i=\frac{d^i}{|{\bf d}|}$ leads to the same result as above with the unique choice of components $b=x, c=y, a=z$ and in that
case the permutation symbol $\epsilon^{abc}=+1$.
\\

{\color{blue} Spin-$\frac{1}{2}$ analogy and Representation(s) with two spheres, Many-body local gauge potential.---} 
On the sphere, the spin-$\frac{1}{2}$ Hamiltonian then implies the general form of the ${\bf d}$-vector
\begin{equation}
\label{d2}
{\bf d} = -2d(\cos\phi\sin\theta,\sin\phi\sin\theta,\cos\theta).
\end{equation}
In general, it is not as simple to acquire a simple understanding of the equivalence between ${\bf d}=(d_x,d_y,d_z)$ and (\ref{d2}).
We present below two representations where the mapping on the sphere will be very useful. Along the line with $k_x=\pi$ linking $X=(\pi,0)$ to $M=(\pi,\pi)$, assuming e.g. that $\Delta\in \mathbb{R}$ with $\varphi=0$, the ${\bf d}$-vector takes the precise form
\begin{equation}
{\bf d} = -(2\Delta \cos\zeta \sin k_y, 2\Delta\sin k_y \sin\zeta, \xi_{k_y}(k_x=\pi,k_y)),
\end{equation}
with $\xi_{k_y}(k_x=\pi,k_y)=-2t\cos k_y -(\mu-2t)$. It will allow for a simple linear correspondence between $\theta$ and $k_y$. The phase $\zeta$ can be linked with the azimuthal angle $\phi$ on the sphere.
For the $p_x+ip_y$ superconductor, we also have the representation with $\Delta=|\Delta|e^{i\varphi}$
\begin{equation}
{\bf d} = -(2\Delta \sin\varphi \sin k_y, 2\Delta\cos\varphi\sin k_y, \xi_{k_y}(k_x=\pi,k_y)),
\end{equation}
showing the possibility to map the angle $\phi$ on the sphere with the superfluid phase $\varphi$. If we redefine $\varphi\rightarrow \frac{\pi}{2}-\varphi$ then we identify $\varphi\leftrightarrow \phi$.
We develop this approach below.
The method from the poles of the sphere will allow us to have an equivalence between a sphere, a cylinder and ellipse geometry... such that this does not require to introduce the precise value of $d$ in this correspondence.
We can also navigate on the line with $k_y=0$ such that
\begin{equation}
{\bf d} = -(2\Delta \cos\varphi \sin k_x, -2\Delta\sin\varphi \sin k_x, \xi_{k_x}(k_x,k_y=0)).
\end{equation}
We have restored a general value of the superfluid phase that will then play the role of a 1D variable on this map and
$\xi_{k_x}(k_x,k_y=0)=-2t\cos k_x-(\mu+2t)$. For this situation, we can find a simple linear transformation between $k_x$ and $\theta$ and similarly between $\varphi$ and $\phi$. 
\\

In this Section, it is also useful to introduce the correspondence between the method on the sphere ({\it see articles [14,18,37-39] in the Letter}) and the introduction of the {\it many-body Chern number} for the BCS theory.  
Absorbing the superfluid phase associated to $\Delta$ onto the product $c^{\dagger}_{\bf k} c^{\dagger}_{-{\bf k}}$, the BCS wavefunction reads
\begin{equation}
|BCS\rangle = \prod_{\bf k}^{\prime} (u_{\bf k}+ v_{\bf k} e^{i\varphi} c^{\dagger}_{\bf k} c^{\dagger}_{-{\bf k}})|0\rangle = \prod_{\bf k} |BCS\rangle_{\bf k},
\end{equation}
where the $\prime$ takes into account the presence of Cooper pairs once. We can then introduce the many-body gauge potential
associated to $|BCS\rangle_{\bf k}$ as
\begin{equation}
A_{\varphi}(k_x,k_y) = (-i)\times\hskip-0.1cm_{\bf k}\langle BCS| \partial_{\varphi} |BCS\rangle_{\bf k} = |v_{\bf k}|^2 = \langle c^{\dagger}_{\bf k} c_{\bf k}\rangle = \frac{1}{2}\left(\langle S_z\rangle +1\right).
\end{equation}
\\

Then, we can navigate on the line along $k_y$ fixing $k_x=\pi$ and $A_{\varphi}=A_{\varphi}(k_y)$. If we suppose that $\mu<4t$ i.e. $\mu$ is e.g. close to half-filling then $A_{\varphi}(\pi,\pi)=0$.
From the mapping on the sphere we also have
\begin{eqnarray}
|v_{k_y}|^2 &=& \frac{1}{2}\left( 1-\frac{\xi_{k_y}}{E_{k_y}}\right) \\ \nonumber
&=& \frac{1}{2}\left(1-\cos\theta\right) = \sin^2\frac{\theta}{2},
\end{eqnarray}
with the correspondence $\xi_{k_y}=-2t\cos k_y -(\mu-2t)$, $\Delta_{k_y}=2e^{i\zeta}\Delta \sin k_y$ and $\theta=k_y\in [0;\pi]$ such that $E_{k_y}=\sqrt{\xi_{k_y}^2 +|\Delta_{k_y}|^2}$.
In this way,
\begin{eqnarray}
|u_{k_y}|^2 &=& \frac{1}{2}\left( 1+\frac{\xi_{k_y}}{E_{k_y}}\right) \\ \nonumber
&=& \frac{1}{2}\left(1+\cos\theta\right) = \cos^2\frac{\theta}{2}.
\end{eqnarray}
Since $\xi_{k_y}(\pi,\pi)>0$ then the $M$ point will be at the {\it north pole} characterized through $\cos\theta_{(\pi,\pi)}=+1$ and for $\mu>0$ we have $\xi_{k_y}(\pi,0)=-\mu$ such that the $X$ point is at {\it south pole}.
These special points indeed satisfy that $|u_{k_y} v_{k_y}|=0$. 
For $\mu=0$ i.e. at the topological transition, since the point $X$ is also classified according to $\cos\theta=0$ then this effectively corresponds to the equatorial plane or to a half Skyrmion. For $\mu<0$, 
the geometrical properties associated to the $X$ and $M$ points are equivalent. 
For the 1D $p_y$-wave Kitaev superconductor, the topological invariant $C_y$ in the article then can be introduced as 
\begin{eqnarray}
C_y = A_{\phi}(\theta_{(\pi,0)})-A_{\phi}(\theta_{(\pi,\pi)}) = A_{\varphi}(k_x=\pi,k_y=0)-A_{\varphi}(k_x=\pi,k_y=\pi).
\end{eqnarray}
{\it The formula from the poles of the sphere can be found in Refs. [14,15,18].}
This is also equivalent to
\begin{equation}
C_y = \frac{1}{2}(\langle S_z(\theta_{(\pi,0)})\rangle - \langle S_z(\theta_{(\pi,\pi)})\rangle).
\end{equation}
The equations take into account that for the Anderson pseudo-spin $A_{\varphi}=\frac{1}{2}\langle S_z\rangle +\frac{1}{2}$ with
$\langle S_z\rangle = -\cos\theta$ such that $C_y=+1$ for $0<\mu<4t$. We have an implicit identification with the lowest-energy eigenstate of the spin-$\frac{1}{2}$ model $H=-{\bf d}\cdot {\bf S}$ with a radial magnetic field such that
where $A_{\phi}$ is directly related to $A_{\varphi}$ modulo a constant, but $C_y$ is gauge invariant. The invariant can be re-written as
\begin{equation}
C_y = \int_{\theta=\theta_{(\pi,\pi)}}^{\theta=\theta_{(\pi,0)}} d\theta \partial_{\theta} A_{\phi}(\theta),
\end{equation}
with the Berry curvature
\begin{equation}
F_{\theta\phi} = \partial_{\theta} A_{\phi}(\theta)= \frac{\sin\theta}{2}
\end{equation}
which can be interpreted in terms of the many-body Chern number as the introduction of a Berry curvature $\partial_{\theta} A_{\varphi}$ such that
\begin{equation}
F_{\theta\varphi} = \partial_{\theta} A_{\varphi}(\theta) = \frac{\sin\theta}{2}.
\end{equation}
We can also verify that due to properties of the coefficients $u_{k_y}$ and $v_{k_y}$, $_{k_y}\langle BCS| \partial_{k_y} |BCS\rangle_{k_y}=0$ in general. 
Since $A_{\varphi}(k_x=\pi,k_y=\pi)=0$ if $\mu<4t$ then the 1D topological invariant $C_y$ also reads $A_{\varphi}(\pi,0)=\langle c^{\dagger}_X c_X\rangle$. This is equal to $+1$ if $\mu>0$ and equal to 0 when $\mu<0$. 
At the transition, the Dirac point is semimetallic associated to a $\frac{1}{2}$ invariant described through $\langle c^{\dagger}_X c_X\rangle=\frac{1}{2}$.
We can formulate a similar analysis for the $C_x$ topological number relating the two points $\Gamma=(0,0)$ and $M=(\pi,0)$.

For the $C_x$ 1D invariant, we can proceed in the same way with the modifications that $C_y\rightarrow C_x$ and $(\pi,\pi)\rightarrow (0,0)$. It is important to mention that when going from $\mu\geq 0$ to $\mu\leq 0$, then
the $d_z$ term flips its sign which means that we can adapt the proof such that $\langle S_z\rangle=\cos\theta$ and $\cos\theta=-\frac{\xi_{k_x}}{E_{k_x}}$ with $\xi_{k_x}=\xi_{\bf k}(k_x,k_y=0)$. In this case, the point $\Gamma=(0,0)$ is at {\it north pole} 
and the point $X=(\pi,0)$ is at {\it south pole}. We obtain then the similar formula
\begin{equation}
\label{EqCx}
C_x = A_{\varphi}(\theta_{(\pi,0)}) - A_{\varphi}(\theta_{(0,0)})=\frac{1}{2}(\langle S_z(\theta_{(\pi,0)})\rangle - \langle S_z(\theta_{(0,0)})\rangle).
\end{equation}
\\

{\color{blue} Local invariant through the sign of kinetic terms and the 1D Pfaffians and through the diagonal in the Brillouin zone.---} In one dimension, associated to the $\mathbb{Z}$ invariant $C_x$ we can also introduce a $\mathbb{Z}_2$ topological number {\it (see Refs. [14,28] in the article)}
\begin{equation}
2C_x^2 -\frac{1}{2}\alpha_x = - \langle S_z(\theta_{(\pi,0)})\rangle \langle S_z(\theta_{(0,0)})\rangle = -\prod_{X,\Gamma} \hbox{sgn}(\xi_{k_x}),
\end{equation}
where 
\begin{equation}
\alpha_x = \langle S_z(\theta_{(\pi,0)})\rangle^2 + \langle S_z(\theta_{(0,0)})\rangle^2.
\end{equation}
The $\mathbb{Z}_2$ invariant refers to the fact that this number is introduced as $\pm 1$ i.e. defined modulo a sign in {\it all the phases} such that $\alpha_x=2$. 
Again, from symmetry, the point $X$ in this analysis equivalently defines a point $Y$.
This invariant is equal to $+1$ within the topological phase with $C_x=-1$ for $-4t<\mu<0$ and to $-1$ in other phases where $C_x=0$. It also relates the pseudo-spin representation with the Pfaffian associated to the 1D Kitaev p-wave superconductor.
This invariant which is defined modulo a sign can also be introduced to characterize the topological transitions at $\mu=-4t$ and $\mu=0$ such that $C_x=-\frac{1}{2}$ and $2C_x^2-\frac{1}{2}\alpha_x=0$. This implies that
$\langle S_z\rangle=0$ at the point where the gap is closing at a topological phase transition that indeed corresponds to a $\frac{1}{2}$ mean local density.

We can then introduce a  $\mathbb{Z}_2$ topological number associated to $C_y$
\begin{equation}
2C_y^2 -\frac{1}{2}\alpha_y = - \langle S_z(\theta_{(\pi,0)})\rangle \langle S_z(\theta_{(\pi,\pi)})\rangle = -\prod_{X,M} \hbox{sgn}(\xi_{k_y}),
\end{equation}
where 
\begin{equation}
\alpha_y = \langle S_z(\theta_{(\pi,0)})\rangle^2 + \langle S_z(\theta_{(\pi,\pi)})\rangle^2.
\end{equation}

Following the method associated to the introduction of these one-dimensional invariants,  we can then write the $\mathbb{Z}$
topological invariant in terms of the $\hbox{sgn}(\xi_{k_x})$ and $\hbox{sgn}(\xi_{k_y})$ as follows. We can equivalently introduce for the present situation the 1D invariants $-|C_x|$ and $|C_y|$, introducing $C=|C_y|-|C_x|$.
In the different phases with $C=0,\pm 1$, we also have $|C_x|=C_x^2$ and $|C_y|=C_y^2$. This allows us to introduce a formula relating $C = |C_y| - |C_x|$ with $\hbox{sgn}(\xi_{\bf k})$ at specific points.
We identify the simplified formula below, that works also at all the topological quantum phase transitions
\begin{equation}
C = |C_y| - |C_x| = \frac{1}{2}\left( \langle S_z(\theta_{(\pi,0)})\rangle \langle S_z(\theta_{(0,0)})\rangle - \langle S_z(\theta_{(\pi,0)})\rangle \langle S_z(\theta_{(\pi,\pi)})\rangle\right).
\end{equation}
This is equivalent to
\begin{equation}
C = -\frac{1}{2}\left(\prod_{X,M} \hbox{sgn}(\xi_{k_y}) - \prod_{\Gamma,X} \hbox{sgn}(\xi_{k_x})\right).
\end{equation}
This is the simplified version of
\begin{equation}
C_y^2-C_x^2 = -\frac{1}{2}\left(\prod_{X,M} \hbox{sgn}(\xi_{k_y}) - \prod_{\Gamma,X} \hbox{sgn}(\xi_{k_x})\right) +\frac{1}{4}(\alpha_y-\alpha_x),
\end{equation}
corresponding then to the difference of 1D $\mathbb{Z}_2$ topological numbers. In this formula,
\begin{eqnarray}
\xi_{k_x} &=& \xi_{k_x}(k_x,k_y=0)=-2t\cos k_x - (\mu+2t) \\ \nonumber
\xi_{k_y} &=& \xi_{k_y}(k_x=\pi,k_y)=-2t\cos k_y - (\mu-2t).
\end{eqnarray}
At $X=(\pi,0)$ this corresponds to
\begin{equation}
\xi_{k_x}(X)=\xi_{k_y}(X)=-\mu.
\end{equation}
The topological invariant $|C_y|-|C_x|$ then corresponds to navigate on a diagonal in the Brillouin zone, such that
\begin{equation}
C = |C_y|-|C_x| = -\hbox{sgn}(\mu)\frac{1}{2}\int_{0}^{\pi} dk_{\parallel} \frac{\partial \langle S_{z}(k_{\parallel})\rangle}{\partial k_{\parallel}} =  -\hbox{sgn}(\mu){\cal W}
\end{equation}
with $k_{\parallel}=\frac{k_x+k_y}{2}$. Then, ${\cal W}$ has also the meaning of a 1D $\mathbb{Z}$ invariant on the sphere where the polar angle is associated to the diagonal direction {\it (see Ref. [14] in the article for a discussion
close to $\mu=-4t$)}. 

Then, this allows us to re-build interesting topological relations on a triangle between $C_x$, $C_y$ and $\hbox{sgn}(\mu){\cal W}$.

\section{Quasiparticles of the Semimetal and Majorana Fermions}

First, we identify the quasiparticles at $\mu=0$ in momentum space forming the semimetal e.g. around the $X=(\pi,0)$ point and then we verify the possibility of a Majorana wavefunction along an edge in the vicinity of $\mu=0$.

We can find the quasiparticle operators associated to the $2\times 2$ matrix in Eq. (\ref{model}) around the $X=(\pi,0)$ point for $\mu=0$. 
In that case, the matrix has only off-diagonal terms such that solutions are superpositions of particle and hole
\begin{equation}
\frac{1}{\sqrt{2}}
\begin{pmatrix}
e^{i\varphi} \frac{\delta k_x+e^{-i\zeta} k_y}{|{\bf k}|} \\
e^{-i\varphi}  \frac{\delta k_x+e^{i\zeta} k_y}{|{\bf k}|} 
\end{pmatrix}.
\end{equation}
At $\mu=0$, the eigenstates are evaluated to linear order in $\delta k_x$ and $k_y$ such that $k_x=\pi-\delta k_x$. The electron and hole refer then to the operators $c_{(\pi,0)}$ and $c^{\dagger}_{(-\pi,0)}$.
Since the Hamiltonian is hermitian this would also accept solutions of the same form with operators $c^{\dagger}_{(\pi,0)}$ and $c_{(-\pi,0)}$. Combining these two solutions, we obtain superpositions which are Majorana fermions.
For the 1D situation, the existence of those Majorana fermions at one point in momentum space also implies a fractional $\frac{1}{2}$ topological invariant {\it (see articles [38,40])}, which agrees with the analysis in the article
$C_x=-\frac{1}{2}$ and $C_y=\frac{1}{2}$ at the topological phase transition at half-filling.
\\

When deviating from $\mu=0$, since we reach the plateau e.g. $C=-1$ or $C=+1$ we should find one Majorana-fermion edge solution. We impose an open boundary condition at $y=0$ and look for zero-energy solutions in the half plane $y>0$. We look
for wave-function solutions of the form 
$$
e^{i qx} e^{-|\lambda| y},
$$
and assume a superposition of a {\it quasihole} and {\it quasiparticle} 
\begin{equation}
\gamma = \int d{\bf r} (u({\bf r}) c({\bf r}) + v({\bf r}) c^{\dagger}({\bf r})).
\end{equation}
Since the $2\times 2$ matrices close to $\Gamma=(0,0)$, $(0,\pi)$ and $(\pi,0)$ are similar we should find identical solutions along the edge. Fourier transforming the matrix, for $\zeta=\frac{\pi}{2}$, we look for Majorana solutions 
\begin{equation}
\left(
\begin{matrix}
\frac{m}{2} & -i\Delta(\partial_x -i\partial_y)\\
-i\Delta^*(\partial_x +i\partial_y) & -\frac{m}{2}
\end{matrix}
\right)
\left(
\begin{matrix}
u({\bf r})\\
v({\bf r})
\end{matrix}
\right) = E\left(
\begin{matrix}
u({\bf r})\\
v({\bf r})
\end{matrix}
\right).
\end{equation}
We make an identification between momentum component $\hbar k_i$ with $i=x,y$ and real space representation $-i\partial_{x_i}$ with $x_i=x$ or $y$.
The eigen-energies are
\begin{equation}
E^2 = \left(\frac{m}{2}\right)^2  + |\Delta|^2(q^2-|\lambda|^2) = |\Delta|^2 q^2.
\end{equation}
This solution satisfies $u({\bf r})=v^*({\bf r})$ i.e. $(u({\bf r}) c({\bf r}))^* = v({\bf r})c^{\dagger}({\bf r})$. 
Assuming a 1D propagating solution along the edge, we can then set $|\lambda|$ such that $|m|^2=4|\lambda|^2 |\Delta|^2$ i.e. $|\lambda|^2\sim R^2$. This ensures the existence of a zero-mode Majorana fermion at the edge when $m\neq 0$.
When $m$ changes of sign e.g. around $\mu=0$ this corresponds to flip the role of $u({\bf r})$ and $v({\bf r})$ associated to a particle-hole transformation. In the sense of Majorana fermions, the Majorana fermion along the edge then turns
in different directions associated to $C_x=-1$ and $C_y=+1$ respectively on both sides of the transition.

\section{Integration and Resolution in Momentum Space}

Here, we address the relevance of the resolution of the measure in momentum space from light or ARPES when integrating on the typical dimensionless ``variable'' ${\cal R}$ around each special point e.g. around the $Y$ point $(\pi,0)$ and 
the $X$ point $(0,\pi)$ in the wavevector space. Below, we study the result on {\it integrated electron density} when selecting ${\cal R}\sim \frac{|\Delta|}{t}$ with $|\Delta|\sim 10K$ and $t\sim 1eV$ i.e. ${\cal R}\sim 10^{-3}$ which is within the experimental capabilities. 
We emphasize here that ${\cal R}$ is dimensionless meaning that we have fixed the lattice spacing to unity and the integration variable is effectively dimensionless. From a physical point of view the superconducting gap $\Delta_{\bf k}$ develops around each gapless special point $\Gamma$, $X$, $Y$ and $M$, which justifies the form of the selected resolution uncertainty. In this Section, we will show that it is indeed possible to resolve the local density of electron with a resolution corresponding then to less of an electron, reproducing the topological phase diagram. In addition, we show that the integrated local density response associated to the physical protocol can then reveal interesting information through its derivative with $\mu$
such as the logarithmic profile of the Van Hove singularity of free fermions on the square lattice defining then the topological phase transition around half-filling.
\\

For special points within the Brillouin zone such that $\Delta_{\bf k}\rightarrow 0$ we have the precise identity
\begin{equation}
\langle c^{\dagger}_{\bf k} c_{\bf k}\rangle = \frac{1}{2} - \frac{1}{2}\hbox{sgn}(\xi_{\bf k}).
\end{equation}
Close to the $(\pi,0)$ or $(0,\pi)$ points, $\hbox{sgn}(\xi_{\bf k})=-\hbox{sgn}(\mu)$. Therefore, when $\mu=0$, $\langle c^{\dagger}_{\bf k} c_{\bf k}\rangle = \frac{1}{2}$ only and when $\mu\neq 0$, we reveal the additional $\hbox{sgn}$ function.
We can also write down the precise mathematical identity introducing the small area $R^2$ in cartesian coordinates such that
\begin{equation}
\langle c^{\dagger}_{X} c_{X}\rangle_{\cal R} = \hbox{lim}_{{\cal R}\rightarrow 0} \frac{1}{\hbox{Area}({\cal R})}\iint d^2{\bf k} \langle c^{\dagger}_{\bf k} c_{\bf k}\rangle.
\end{equation}
The question is then to estimate what means ${\cal R}\rightarrow 0$ to reproduce the local electron density at a special point in momentum space? 
For this purpose, we analyse the functions $\xi_{\bf k}$ and $E_{\bf k}$ that enter in 
\begin{equation}
\langle c^{\dagger}_{\bf k} c_{\bf k}\rangle = \frac{1}{2}\left(1-\frac{\xi_{\bf k}}{E_{\bf k}}\right),
\end{equation}
close to the $Y$ point $(0,\pi)$ developing 
$k_x=\delta k_x$ and $\delta k_y=\pi-k_y$ with $(\delta k_x,\delta k_y)\ll 1$:
\begin{equation}
\xi_{\bf k}=\left(t(\delta k_x^2-\delta k_y^2)-\mu\right)
\end{equation}
\begin{equation}
E_{\bf k}=\sqrt{4|\Delta|^2(\delta k_x^2+\delta k_y^2) + (t(\delta k_x^2-\delta k_y^2)-\mu)^2}.
\label{Ek}
\end{equation}
For simplicity, we address the situation of a positive value of $\mu$, but the situation of negative $\mu$-values can be symmetrically addressed.
\\

First, selecting a value of $\mu$ such that $\mu\gg t{\cal R}^2$ then 
\begin{equation}
\frac{1}{{\cal R}^2}\iint_0^{{\cal R}^2} d^2{\bf k} \left(\frac{1}{2} - \frac{1}{2}\hbox{sgn}(\xi_{\bf k})\right) \sim \frac{1}{2}+\frac{1}{2}\hbox{sgn}(\mu) = \langle c^{\dagger}_X c_X\rangle = \langle c^{\dagger}_Y c_Y\rangle.
\label{densityintegration}
\end{equation}
In this way, on both sides of the phase transition, we do probe a nice {\it quantized plateau} for the electron density at $X=(0,\pi)$ and $Y=(\pi,0)$. Summing the two responses, the $\hbox{sgn}(\mu)$ then measures either the topological
invariant $-1$ or $+1$ around the topological phase transition. 
\\

We are now questioning about the physical response when approaching the Van Hove situation corresponding to fix $\mu\sim \frac{2|\Delta|^2}{t}$. 
It should be emphasized that this already corresponds to a {\it very small value of} $\mu$ in absolute value such that we are approaching the transition. Our goal is precisely to show that when taking into account the resolution of the
protocol in momentum space, this is precisely the (small) chemical potential value(s) around $\mu=0$ showing the transition or passage from density $1$ to density $0$ associated to a logarithm profile of the derivative as a function of $\mu$ instead of the Heaviside step function
illustrating the behavior of the topological invariant at $T=0$.

The dominant terms to the integrated electron density read
\begin{equation}
\iint d^2{\bf k} \frac{1}{2}\left(1+\frac{\mu-t\delta k_x^2}{\sqrt{(4|\Delta|^2+2\mu t)\delta k_y^2 +\mu^2}}\right).
\label{integrallog}
\end{equation}
Around half-filling, we have simplified the expression at the numerator 
dropping the the term with $t\delta k_y^2$. This term would lead to the integral
\begin{equation}
\int \frac{x^2}{\sqrt{a x^2 + b}} dx = \frac{1}{2}\left(\frac{x\sqrt{ax^2+b}}{a}-\frac{b}{a^{\frac{3}{2}}}\tanh^{-1}\left(\frac{\sqrt{a}x}{\sqrt{ax^2+b}}\right)\right),
\end{equation}
where $a=(4|\Delta|^2+2\mu t)$, $b=\mu^2$ and $x=\delta k_y$. Since $b\propto \mu^2$ the second term and its derivative will be very small compared to the logarithm response found below. The first term is also less dominant when $x$ is small.
Introducing $\alpha=\sqrt{4|\Delta|^2+2\mu t}$, then the dominant response is
\begin{equation}
\iint_0^{{\cal R}^2}d^2{\bf k} \langle c^{\dagger}_{\bf k} c_{\bf k}\rangle \sim \frac{1}{2}\hbox{Area} + \left(\frac{\mu}{2}{\cal R}-\frac{t}{2}\frac{{\cal R}^3}{3}\right)\left(\frac{1}{\alpha}\ln\left(\sqrt{{\cal R}^2+\frac{\mu^2}{\alpha^2}}+{\cal R}\right)-\frac{1}{\alpha}\ln\frac{|\mu|}{\alpha}\right).
\end{equation}
If we normalize the result to a small area ${\cal R}^2$ then the result corresponds to
\begin{equation}
\frac{1}{{\cal R}^2}\iint_0^{{\cal R}^2}d^2{\bf k} \langle c^{\dagger}_{\bf k} c_{\bf k}\rangle \sim \frac{1}{2} +\frac{\mu}{2{\cal R}}\left(\frac{1}{\alpha}\ln\left(\sqrt{{\cal R}^2+\frac{\mu^2}{\alpha^2}}+{\cal R}\right)-\frac{1}{\alpha}\ln\frac{|\mu|}{\alpha}\right).
\label{mu}
\end{equation}
If we fix ${\cal R}\ll \frac{|\mu|}{\alpha}$ then the result is almost $\frac{1}{2}$ already. If we increase ${\cal R}\gg\frac{|\mu|}{\alpha}$ in the first logarithmic term then 
\begin{equation}
\langle c^{\dagger}_X c_X\rangle = \langle c^{\dagger}_Y c_Y\rangle \sim \frac{1}{{\cal R}^2}\iint_0^{{\cal R}^2}d^2{\bf k} \langle c^{\dagger}_{\bf k} c_{\bf k}\rangle \sim \frac{1}{2} - \frac{\mu}{2 \alpha {\cal R}}\ln\frac{|\mu|}{{\cal R}\alpha}.
\label{Xdensity}
\end{equation}
The correction to $\frac{1}{2}$ for the local density is yet very small such that we verify that setting $\mu=\frac{2|\Delta|^2}{t}\rightarrow 0$ corresponds to reach the vicinity of the topological phase transition when integrating on a small area around the $X$ point (or $Y$ point). Then, we verify that the jump in the topological number indeed corresponds in this way to a
 {\it logarithmic profile} of the derivative $\frac{\partial \langle c^{\dagger}_X c_X\rangle}{\partial\mu}$ at the (small) Van Hove filling, which is an interesting remark for physical and practical applications.
\\

It is important to mention that when including a off-diagonal hopping term such that 
\begin{equation}
\xi_{\bf k}=-2t\cos k_x -2t\cos k_y - t' \cos k_x \cos k_y-\mu
\end{equation}
then to satisfy $\xi_{\bf k}=0$ at an $X$ or $Y$ point this implies to slightly relocate the chemical potential such that $\mu=t'$. This shifts accordingly the transition point at the filling factor $\mu=t'$. The resulting topological transition at $\mu=t'$ has the same physical properties as above such that 
\begin{equation}
\langle c^{\dagger}_X c_X\rangle = 
\langle c^{\dagger}_Y c_Y\rangle = \frac{1}{2}. 
\end{equation}
The Van Hove fillings then correspond to the two specific values $\mu_{eff}=\mu-t'=\frac{2|\Delta|^2}{t-\frac{t'}{2}}$ and 
$\mu_{eff}=\mu-t'=-\frac{2|\Delta|^2}{t+\frac{t'}{2}}$; as long as $t'\ll t$
the physics remains almost identical as the one described above through the logarithmic derivative of the density of states. In that case, around half-filling the spread of the Berry curvature would be on a radius ${\cal R}\sim \frac{t'}{\Delta}$.

\section{Correlation Functions, Local Marker and Electron Density in Real Space}

\begin{figure}[t]
\hskip 2cm \includegraphics[width=12cm]{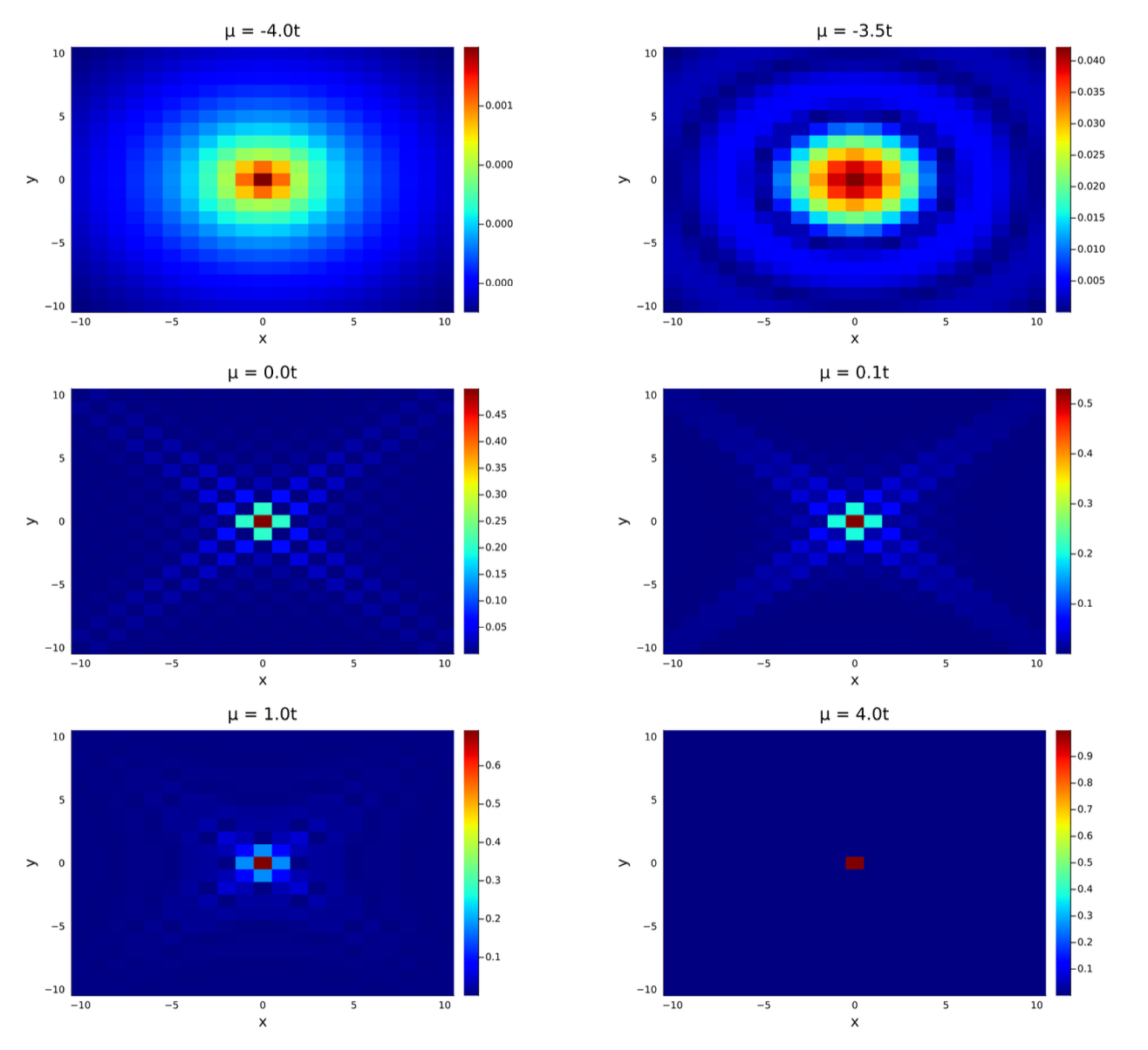}
\caption{2D Representation of correlation functions in real space. It is important to emphasize that the scale on the vertical axis measuring also the importance of the local electron density
for $\Delta_x=\Delta_y=0$ is different around $\mu=-4t$ and $\mu=4t$. We take $\Delta=0.1$ in units of $t$ for all the figures.}
\label{drawing}
\end{figure}

Here, we study observables in real space associated to the topological local marker such as correlation functions and electron density in real space.
The local marker for the $p_x+ip_y$-wave superconductor reads
\begin{equation}
\label{invariantequation}
C = C_x+C_y = -\langle c^{\dagger}_{\Gamma} c_{\Gamma}\rangle +2\langle c^{\dagger}_X c_X\rangle -\langle c^{\dagger}_M c_M\rangle.
\end{equation}
This invariant involves three specific points e.g. $(0,0)$, $(\pi,0)$ and $(\pi,\pi)$. 
From Fourier transform
\begin{eqnarray}
\label{realspace}
C = \frac{1}{N}\sum_{i,j} \langle c^{\dagger}_i c_j\rangle (- 1 +2 e^{i {\bf k}_X \cdot ({\bf r}_i - {\bf r}_j)} - e^{i {\bf k}_M \cdot ({\bf r}_i - {\bf r}_j)}).
\end{eqnarray}
with $N$ being the number of sites.
Introducing $({\bf r}_i - {\bf r}_j)=(\Delta_x,\Delta_y)$, the correlation functions read
\begin{equation}
\langle c^{\dagger}_i c_j\rangle = \frac{1}{\pi^2}\int_0^{\pi} dk_x \cos(k_x\Delta_x)\int_0^{\pi} dk_y \cos(k_y \Delta_y) \frac{1}{2}\left(1-\frac{\xi_{\bf k}}{E_{\bf k}}\right).
\end{equation}
It takes an equivalent form
\begin{equation}
\langle c^{\dagger}_i c_j\rangle = \frac{1}{\pi^2}\int_0^{\pi} dk_x \cos(k_x \Delta_x) \int_0^{\pi} dk_y\cos(k_y\Delta_y) (-)\frac{1}{2}\frac{\xi_{\bf k}}{E_{\bf k}}.
\end{equation}
In Fig. \ref{drawing}, we show all correlation functions in real space as a function of $\mu$. 

At $\mu=0$ when $t'=0$, we can again apply particle-hole symmetry transformation. In that case, it gives rise to the identity along the diagonal $\Delta x=\Delta y$
\begin{equation}
\langle c^{\dagger}_i c_j \rangle = - \langle c^{\dagger}_i c_j \rangle.
\end{equation}
This represents a signature of the phase transition at $\mu=0$. 
When including $t'$, a similar result is obtained for the generalized filling for which $\xi_{\bf k}=0$. The analysis of the correlation functions in different directions at $\mu=0$ 
will be addressed later.

 \begin{figure}[t]
\hskip 3cm \includegraphics[width=11cm]{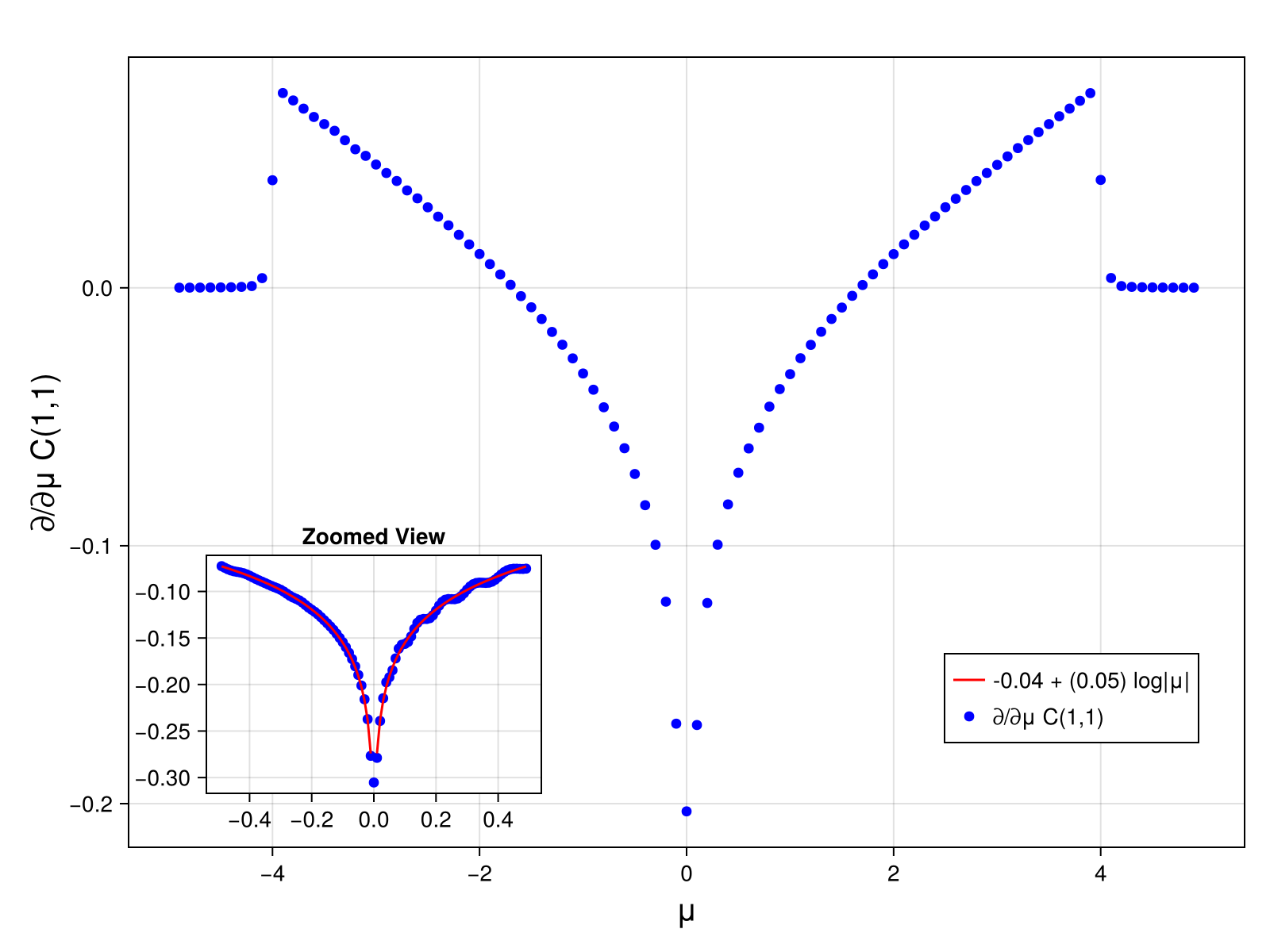}
\caption{Derivative of the correlation function with $\Delta_x=\Delta_y=1$.}
\label{DerivativeC11}
\end{figure}

\begin{figure}[t]
\hskip 2cm \includegraphics[width=11cm]{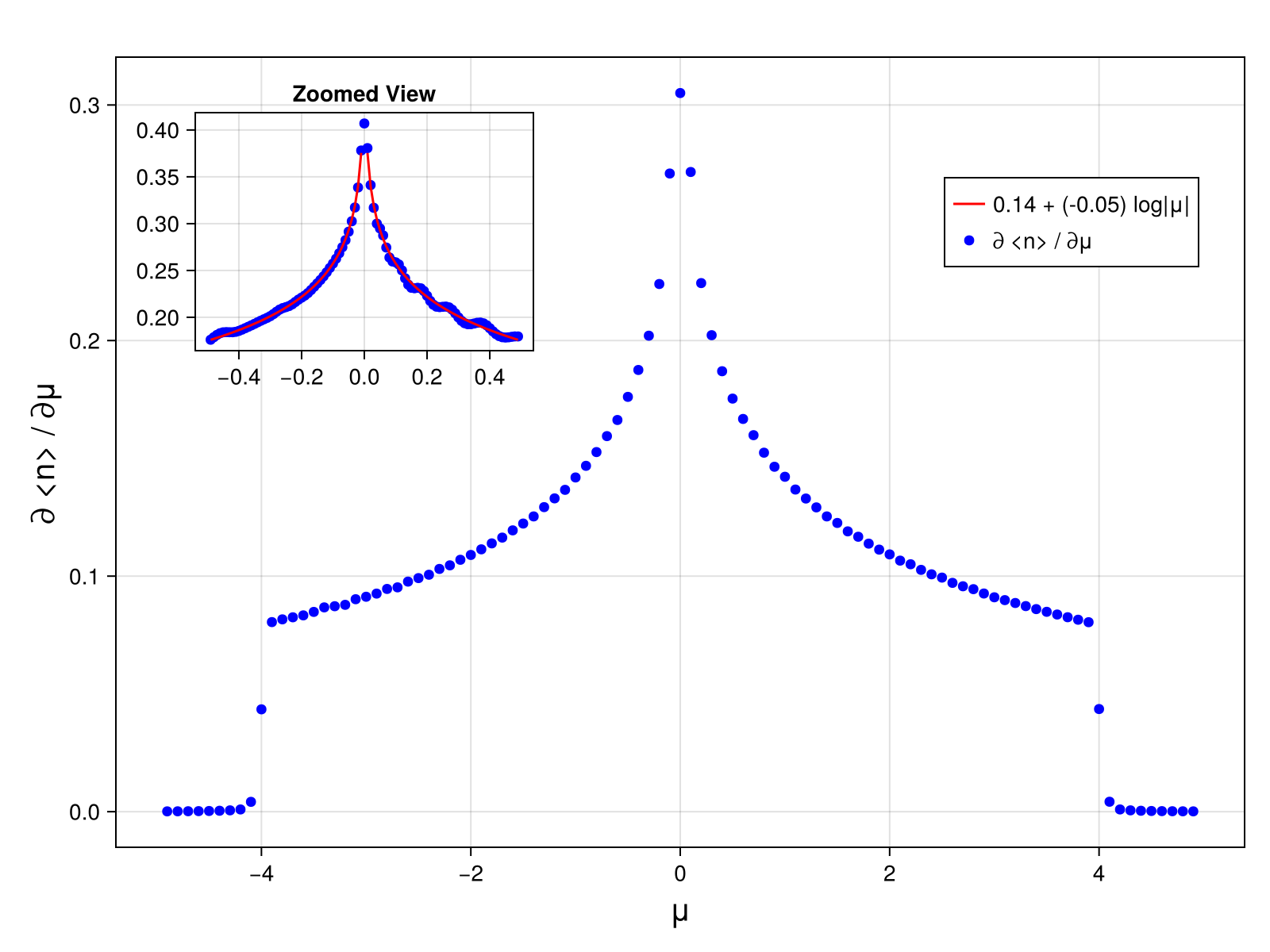}
\caption{Derivative of the local electron density in real space.}
\label{DerivativeDensity}
\end{figure}

For $\mu\neq 0$ (for simplicity we assume $t'=0$ but the results are similar when including $t'$ modulo the redefinitions of $m$, $\mu_{eff}$ in Eq. (\ref{log})), we can then apply the results above
\begin{equation}
\langle c^{\dagger}_i c_j\rangle = \frac{1}{\pi^2}\int_0^{\pi} dk_x \cos(k_x\Delta_x)\int_0^{\pi} dk_y \cos(k_y \Delta_y) \langle c^{\dagger}_{\bf k} c_{\bf k}\rangle_{\mu}.
\end{equation}
Close to the $(\pi,0)$ or $(0,\pi)$ point, if we set the correlation function $C(1,1)$ corresponding to $\Delta x = \Delta y=1$ then
we have $\cos(k_x \Delta_x)\cos(k_y \Delta_y)=-1$ such that the logarithmic profile of the derivative with $\mu$ is negative, as observed numerically in Fig. \ref{DerivativeC11}.
We also report a positive derivative with $\mu$ with a similar logarithmic profile for the local density in real space when setting $\Delta_x=\Delta_y=0$; see Fig. \ref{DerivativeDensity}. 
For the correlation functions and local density in real space, the logarithmic profile of the derivative develops on a relative large phase space of $\mu$.
We numerically report that the prefactors do not depend on the ratio $\frac{\Delta}{t}$. 
This maybe understood as follows from the results close to the Van Hove filling. The main result comes from a radius ${\cal R}$ from a point $X$ such that from Eq. (\ref{Xdensity}), we obtain
\begin{equation}
\frac{\partial}{\partial \mu} \langle c^{\dagger}_i c_j\rangle \sim - \frac{2}{\pi^2}\frac{{\cal R}}{2\alpha}\ln\frac{|\mu|}{{\cal R}\alpha}.
\label{log}
\end{equation}
When we set $\mu\sim \frac{2|\Delta|^2}{t}$ then $\alpha\sim \sqrt{8}\Delta$ which means that $\frac{{\cal R}}{\alpha}$ is independent of $\Delta$. The additional factor $2$
takes into account that we have two $X$-points at $(0,\pi)$ and $(\pi,0)$, i.e. equivalent $X$ and $Y$ points, within the area of integration.
Then, the prefactor is of the order of $\sim -0.1$ in units of $\frac{1}{t}$ which is in the correct range of magnitude. 

\vskip 0.5cm
It is then interesting to address the situation when $\mu\rightarrow -4t$ close to the $\Gamma=(0,0)$ point. 
We can simplify $\xi_{\bf k}$ as
\begin{equation}
\xi_{\bf k} = 4t\sin^2\frac{k_x}{2} + 4t\sin^2\frac{k_y}{2}-\delta \mu,
\end{equation}
where $\mu=-4t+\delta\mu$ measures the deviation from $\mu=-4t$. The numerator then can be developed as
\begin{equation}
\xi_{\bf k} \sim t(k_x^2 +k_y^2)-\delta\mu
\end{equation}
and we verify the rotationally invariant form for the $p_x+ip_y$ superconductor (the system is almost empty and we develop the spectrum around $(0,0)$). We can also simplify
\begin{equation}
E_{\bf k} \sim \sqrt{(-2\delta\mu t+4\Delta^2)(k_x^2+k_y^2)+\delta\mu^2}.
\end{equation}
When $\delta\mu<0$, the band occupancy is reduced to the point itself such that $k_x=k_y=0$. In that case, we find that the correlation functions go to zero rapidly.
When $\delta\mu > 0$,
\begin{equation}
\langle c^{\dagger}_i c_j \rangle = \frac{1}{\pi^2}\int_0^{\pi} dk_x \cos(k_x \Delta_x) \int_0^{\pi} dk_y \cos(k_y\Delta_y) (-) \frac{1}{2}\frac{-\delta\mu + t(k_x^2 +k_y^2)}{\sqrt{(-2\delta\mu t+4|\Delta|^2)(k_x^2+k_y^2)+\delta\mu^2}}.
\end{equation}
When $\delta\mu\rightarrow 0^+$, we can simplify the formula as
\begin{equation}
\langle c^{\dagger}_i c_j\rangle =-\frac{1}{\pi^2}\frac{t}{2|\Delta|}\int_0^{\pi} dk_x \cos(k_x \Delta_x) \int_0^{\pi} dk_y\cos(k_y \Delta_y) 
\sqrt{k_x^2+k_y^2}.
\end{equation}
Around $\mu=-4t$ the energy spectrum is {\it rotationally invariant} around $(0,0)$ and similarly for correlation functions in real space; see Fig. \ref{drawing}. 
In polar coordinates $k_x=|{\bf k}|\cos\varphi$ and $k_y=|{\bf k}|\sin\varphi$ through the Bessel function of the first kind:
\begin{equation}
\int_{0}^{2\pi} d\varphi e^{\pm i |{\bf k}|r \cos \varphi} e^{\pm i |{\bf k}| r\sin \varphi} = 2\pi J_0(\sqrt{2}|{\bf k}| r),
\end{equation}
with $r=\Delta_x=\Delta_y$. When developing the cosine and sine functions, the four terms give the same Bessel function (modulo a change of variables $\varphi\rightarrow -\varphi$).
Therefore, when $\delta\mu \rightarrow 0$,
\begin{equation}
\langle c^{\dagger}_i c_j\rangle \sim -\frac{1}{\pi^2}\frac{t}{2|\Delta|}\int_0^{\pi} dk k^2 2\pi J_0(k\sqrt{2}r).
\end{equation}
Then, we do the change of variables $k' = |{\bf k}| r=kr$ such that
\begin{equation}
\langle c^{\dagger}_i c_j\rangle \sim - \frac{1}{\pi^2}\frac{t}{2|\Delta|}\int_0^{\pi r} \frac{dk'}{r^3} 2\pi J_0(\sqrt{2} k') k'^2.
\end{equation}

Then, we obtain the integral
\begin{equation}
\frac{1}{r^3}\int_0^{r\pi} dk' k'^2 J_0(\sqrt{2}k'),
\end{equation}
such that at small $r$ it behaves as $\frac{\pi^3}{3}-{\cal O}(r^2)$ and at large $r$ the dominant scaling is $1/\sqrt{r}$ from the power-law behavior of the Bessel function. 
Numerically, we also report an exponential decay accompanying the power-law behavior for $\mu=-4t+\epsilon$ with $\epsilon>0$ and $\epsilon\rightarrow 0$. 
The correlation functions are discontinuous at the transition $\mu=-4t$ revealing the strong-paired to weak-paired phase transition.
\\

For $\mu=0$, analytically and numerically we do identify a pure algebraic power-law decay at long distances.
This emphasizes that the region close to half-filling indeed corresponds to a weaked-paired phase.
The long-range profile of correlation functions at $\mu=0$ with an infinite correlation length is visible from certain directions in Fig. \ref{drawing} while we emphasize on the localization aspect along the specific line $\Delta_x=\Delta_y$ which also
marks the phase transition at $\mu=0$.

\vskip 0.5cm The local density response with $\Delta_x=\Delta_y=0$ becomes dominant when increasing the chemical potential; see central dot response in Figs. \ref{drawing}.
However, it is then important to mention that the red central dot in  Fig. \ref{drawing} 
does not participate in the local marker in Eq. (\ref{realspace}) if we set $\Delta_x=\Delta_y=0$.
Therefore, to reproduce $C=+1$ in the vicinity of the transition e.g. when $\mu=3.9t$ then this requires to take into account the correlation functions with very weak intensity in Fig. \ref{drawing}.
This results in the Figure in the article. We also find numerically that if we fit the long-range decay of the correlation functions close to the transition at $\mu=4t$, e.g. for $\mu=4t-\epsilon$ with $\epsilon>0$ and $\epsilon\rightarrow 0$, then 
we also report a power-law decay in $\frac{1}{\sqrt{r}}$, but with a multiplying exponential factor. In Figs. \ref{DerivativeC11} and \ref{DerivativeDensity}, we also reveal that the derivatives of different observables are symmetric around $\mu=-4t$ and $\mu=+4t$
and reveal a jump at $\mu=-4t$ and $\mu=+4t$ which is then a signature of those topological transitions.


\section{Robustness of Local Marker Towards Lifetime of Quasiparticles and Frequency Phase Space}

We address here the robustness of the local topological marker through the introduction of a life-time for the quasiparticles, that can result from interaction effects, and we also address the phase space
in frequency. Suppose we introduce a life-time for the Bogoliubov de Gennes quasiparticles. The retarded electron Green's function  then takes the form
\begin{equation}
G_{\bf k}(\omega) = \frac{|u_{\bf k}|^2}{\omega-E_{\bf k}+i\Gamma} + \frac{|v_{\bf k}|^2}{\omega+E_{\bf k}+i\Gamma}.
\end{equation}
The electron spectral function then reads
\begin{equation}
A({\bf k},\omega) = \frac{\Gamma}{\pi}\left(\frac{|u_{\bf k}|^2}{(\omega-E_{\bf k})^2 +\Gamma^2} + \frac{|v_{\bf k}|^2}{(\omega+E_{\bf k})^2 +\Gamma^2}\right).
\end{equation}
\begin{equation}
\langle c^{\dagger}_{\bf k} c_{\bf k}\rangle =\ \frac{\Gamma}{\pi}\int d\omega n_F(\omega)\left(\frac{|u_{\bf k}|^2}{(\omega-E_{\bf k})^2 +\Gamma^2} + \frac{|v_{\bf k}|^2}{(\omega+E_{\bf k})^2 +\Gamma^2}\right).
\end{equation}
Then, we suppose that the lifetime is long such that $\Gamma$ is small and we can then simplify $n_F(\omega)$ from $\omega=\pm E_{\bf k}$. Then, we can verify the proof that
\begin{equation}
\langle c^{\dagger}_{\bf k} c_{\bf k}\rangle = \frac{\Gamma}{2\pi}\int_{-\infty}^{+\infty} d\omega \left(n_F(E_{\bf k}) \frac{1}{(\omega-E_{\bf k})^2 +\Gamma^2} + n_F(-E_{\bf k}) \frac{1}{(\omega+E_{\bf k})^2 +\Gamma^2}\right)
\end{equation}
\begin{equation}
+  \frac{\Gamma}{2\pi}\int_{-\infty}^{+\infty} d\omega\hbox{sgn}(\xi_{\bf k}) \left(n_F(E_{\bf k}) \frac{1}{(\omega-E_{\bf k})^2 +\Gamma^2} - n_F(-E_{\bf k}) \frac{1}{(\omega+E_{\bf k})^2 +\Gamma^2}\right).
\end{equation}
We can integrate the integrals in frequency such that
\begin{equation}
\int_{-\infty}^{+\infty} \frac{1}{x^2+\Gamma^2} dx = \frac{\pi}{\Gamma}.
\end{equation}
Then, the lifetime simplifies and we verify the preceding result for $\Gamma\rightarrow 0$:
\begin{equation}
\langle c^{\dagger}_{\bf k} c_{\bf k}\rangle = \frac{1}{2}(n_F(E_{\bf k}) + n_F(-E_{{\bf k}})) - \frac{1}{2}\hbox{sgn}(\xi_{\bf k})(n_F(-E_{\bf k})-n_F(E_{{\bf k}}))
\end{equation}
\begin{equation}
\langle c^{\dagger}_{\bf k} c_{\bf k}\rangle = \frac{1}{2} - \frac{1}{2}\hbox{sgn}(\xi_{\bf k})\tanh\left(\frac{\beta E_{\bf k}}{2}\right).
\end{equation}
\\

We can also discuss the phase space in $\omega$: In that case, the integrals read
\begin{equation}
\int_{-\lambda}^{\lambda} \frac{dx}{x^2+\Gamma^2} = \frac{2}{\Gamma}\hbox{arctan}\frac{\lambda}{\Gamma}.
\end{equation}
Therefore, as long as the phase space in $\omega$ around the pole satisfies $\lambda\gg \Gamma$, we find the same result.

\end{document}